\documentclass[aps,pre,notitlepage,reprint,onecolumn,showpacs,superscriptaddress]{revtex4-1}
\usepackage{epsfig}
\usepackage{natbib}
\usepackage{ulem}
\usepackage{xspace}
\usepackage{xcolor}
\usepackage{graphicx}
\usepackage{amssymb}
\usepackage{cancel}
\usepackage{ulem}
\usepackage{lineno}
\usepackage{amsmath}
\usepackage{stackrel}
\usepackage{appendix}
\usepackage{url}
 \usepackage{enumitem}
 \usepackage{hyperref}
\usepackage{xcolor}
\usepackage{array}
\usepackage{booktabs}
\usepackage{multirow}
\usepackage{float}

\hypersetup{
	colorlinks=true,
	linkcolor=blue,
	citecolor=blue,
	urlcolor=blue
}

\begin{document}
\title{Extreme Values of Infinite-Measure Processes}

\author{Talia Baravi}
\affiliation{Department of Physics, Institute of Nanotechnology and Advanced Materials, Bar-Ilan University, Ramat-Gan 52900, Israel}

\author{Eli Barkai}
\affiliation{Department of Physics, Institute of Nanotechnology and Advanced Materials, Bar-Ilan University, Ramat-Gan 52900, Israel}  

\begin{abstract} 
We study the statistics of the maximum and minimum of a set of $N$ random variables whose dynamical and statistical properties fall within the scope of infinite ergodic theory. These non-stationary yet recurrent systems are described, in the long-time limit, by a non-normalizable infinite invariant density. Extreme events in such systems emerge in a joint limit where the observation time $t$ is long and the number of variables $N$ is large. We show that the resulting extreme value statistics are controlled by the return exponent $\alpha$ and the infinite invariant measure, and therefore depart from the classical Fréchet, Gumbel, and Weibull universality classes. We illustrate the theory for weakly chaotic intermittent maps, overdamped diffusion in an asymptotically flat potential, and a stochastic model of sub-recoil laser cooling, and show how measurements of extremes can be used to infer the infinite-density structure.
\end{abstract}

\maketitle

\section{Introduction}
Extreme-value statistics provides a natural language for quantifying {rare events}, the largest fluctuation, the earliest arrival, the longest waiting time, or the most extreme excursion observed within a finite large sample of size $N$. Beyond its classical roots in meteorology and risk analysis, extreme-value theory (EVT) has become a standard tool across physics, biology, and complex systems, where extremes often control performance, failure, or optimality \cite{fisher1928limiting,Gnedenko1943,Gumbel1958,anderson1984extremes,embrechts1999modelling,fortin2015applications,schuss2019redundancy,grebenkov2026fastest,lawley2023slowest,jenkinson1955frequency,majumdar2002extreme,ellettari2025rare,lawley2024competition,eliazar2019gumbel,tung2025passage}. In its canonical form, EVT states that the extremum of many independent and identically distributed  (i.i.d.) variables converges, after an appropriate rescaling, to one of three universality classes (Gumbel, Fréchet, Weibull), determined by the tail of the parent distribution \cite{fisher1928limiting,Gnedenko1943,Gumbel1958}. This universality reduces a potentially complicated parent distribution to a small set of limit laws and parameters.

However, two aspects that arise routinely in statistical physics push EVT beyond its textbook setting. First, many physically relevant extremes are formed in {correlated} sequences, where clustering of rare events and memory effects can invalidate the classical convergence mechanism; moreover, finite-size or finite-sample effects can enforce pre-asymptotic, non-classical extreme statistics \cite{majumdar2020extreme,hall1979rate,oshanin2013anomalous,zarfaty2022discrete,mikosch2020gumbel,anderson1984extremes,samorodnitsky2004extreme,biroli2023extreme}. 
Second, even when the sampled variables are independent and identically distributed, the parent distribution may still depend on a large control parameter (for example, the observation time). In that case, extreme-value statistics becomes meaningful in a joint limit where the number of samples $N$ grows together with that control parameter. This type of thermodynamic scaling for extremes was demonstrated recently for extreme first-passage times when system size is made large \cite{baravi2025thermodynamic}. Here we adopt the same viewpoint in a broader class of stochastic and deterministic recurrent processes governed by infinite ergodic theory: we take both the measurement time $t$ and the sample size $N$ to be large, and we show that a nontrivial extreme-value limit emerges only under an appropriate joint scaling controlled by the return exponent $\alpha$.

Many stochastic and deterministic dynamical systems of interest do not possess a normalizable invariant density. In many cases, this behavior is tied to a diverging intrinsic time scale: the processes we consider are recurrent, but the mean return time to a reference region diverges (we define this precisely below). As a result, long-time statistics is organized not by a stationary probability density, but by the structures described by infinite ergodic theory. The limiting scaling function, called the infinite invariant density, is non-normalizable and controls long-time observables and rare fluctuations. In deterministic dynamics this scenario appears, for example, in weakly chaotic (intermittent) maps with marginal fixed points (e.g. Pomeau–Manneville-type maps) \cite{pomeau1980intermittent,thaler1983transformations,thaler2000asymptotics} and in related continuous-time deterministic models \cite{meyer2017infinite}. In stochastic physics, closely related infinite-density behavior arises in heavy-tailed trapping models (e.g. Bouchaud-type trap models), in multiplicative-noise processes, in diffusion through inhomogeneous media, and in settings such as laser-cooling dynamics \cite{giordano2026generalized,akimoto2022infinite,barkai2023ergodic,rebenshtok2014infinite,aghion2020infinite,barkai2022gas,giordano2023infinite,barkai2021transitions,vezzani2019single,akimoto2008generalized}.

Infinite ergodic theory provides the organizing framework for this regime. When the long-time state is described by a non-normalizable invariant density, the usual ergodic picture breaks down: even at long times, time-averaged observables do not settle to a single deterministic value, but remain broadly distributed from trajectory to trajectory. Nevertheless, after a suitable rescaling, these time averages converge in distribution to limit laws such as Darling–Kac distributions \cite{radice2020statistics,aaronson1997introduction,thaler2006distributional,akimoto2015distributional}. Related aging phenomenology appears in trapping models of glassy dynamics. In Bouchaud’s trap model, the low-temperature phase is accompanied by diverging time scales and a non-normalizable equilibrium weight (often phrased as a diverging partition function), leading to persistent trajectory-to-trajectory fluctuations and nonstationary occupation statistics \cite{bouchaud1992weak}.

The goal of this paper is to demonstrate how a non-classical extreme value theory emerges in systems governed by infinite ergodic theory. In these systems, both the long-time limit and the large-$N$ limit arise naturally. Taking this joint limit carefully yields a new extreme value framework, which connects extreme statistics to the infinite invariant density and the return exponent $\alpha$.
This connection is universal in the sense that it applies across a broad class of models. At the same time, because the infinite invariant density is system-specific (much like a normalized invariant density in standard settings), some microscopic details persist and shape the extreme statistics. We therefore study three representative examples: weakly chaotic nonlinear dynamical systems, Langevin diffusion in a non-binding force field, and a model of  laser cooling to clarify how these ingredients control the behavior of extremes.

The paper is organized as follows. In Section II we introduce the infinite-invariant-density framework and derive the joint limit of large $t$ and large $N$, obtaining the limiting laws for maxima/minima in terms of the integrated infinite density. We then illustrate these predictions in three representative settings: weakly chaotic intermittent maps (Section III.a), overdamped diffusion in an asymptotically flat potential (Section III.B), and a renewal model of sub-recoil laser cooling (Section III.C). Section IV concludes with a discussion of the physical implications, limitations, and possible extensions.

\section{Infinite invariant density and extreme value statistics}\label{eq:sec1}
We consider a dynamical process $x(t)$, either stochastic or deterministic, which may represent the coordinate of a particle in one dimension. The system is described statistically by the probability density $p(x,t)$, defined over an ensemble of trajectories. In many scenarios, one focuses on the long-time limit in which $p(x,t)$ approaches a steady state. Yet in many other situations, the long time limit is described by a non-equilibrium state whose spatial shape is time independent, yet whose total weight diverges. In such cases, the conventional steady-state framework breaks down, and the long-time behavior is encoded in an infinite invariant density: a non-normalizable function that replaces the steady-state distribution as the central statistical object. Importantly, the dynamics we consider is recurrent: trajectories return to the relevant region infinitely often with probability one. Thus, while the system is non-ergodic in the usual sense, the repeated visitation of the physically relevant region of phase space is still guaranteed. Concretely, for a suitable observable $x$ with probability density $p(x,t)$, the infinite invariant density $\mathcal{I}(x)$ is defined through the limit
\begin{equation}\label{eq:infden_def}
\mathcal{I}(x)=\lim_{\substack{t\rightarrow \infty}}\mathcal{N}\,
t^{1-\alpha}\,p(x, t) \, , 
\end{equation}
where $x$ is fixed and does not scale with $t$, see below. Here the constant $\mathcal{N}$ is model-specific and will be given later, and $0<\alpha<1$ is the return (persistence) exponent, defined by the tail of the return times distribution (see the model sections below). $\mathcal{I}(x)$ in Eq.\,(\ref{eq:infden_def}) cannot be normalized, namely $\int_{{\cal S}} \mathcal{I}(x)\,dx\rightarrow\infty$ where ${\cal S}$ is the support or domain of $x$, and further this function is invariant under time changes and uniquely determined up to a multiplicative constant. As we discuss below, this infinite invariant density governs long-time statistics, controls rare events, and determines measurable observables.

We consider non-negative random variables $x \geq 0$ and distinguish between two cases:

\begin{equation}\label{eq:clasess}
\mathcal{I}(x) \propto  
\begin{cases} 
x^{-\beta}, & x \rightarrow 0^+ \, \, (\text{Case 1}), \\
x^{-\beta}, & x \rightarrow \infty\, \, (\text{Case 2})\, .
\end{cases}
\end{equation}
We denote by $\beta$ the power-law exponent that characterizes the non-normalizable behavior of the infinite invariant density $\mathcal{I}(x)$. In Case 1 we have $\beta\geq 1$, this occurs for example, for the Pomeau–Manneville map, and in that example we later identify $\beta=1/\alpha$. In Case 2, the non-normalizability is due to the large-$x$ behavior and $\beta \le 1$, see the example of Langevin dynamics in an asymptotically flat potential below. We will later explain the physical meaning of $\alpha$ and $\beta$ for various models, and relate it to features of the dynamics. Formally, $\mathcal{I}(x)$ is a stationary solution of the relevant evolution operator (Perron–Frobenius for maps, Fokker–Planck for stochastic dynamics), but it is non-normalizable, as mentioned, hence it cannot represent a steady-state probability measure.

\subsection{Maximum statistics controlled by $\mathcal{I}(x)$ (case 1)}\label{subsec:ev}
In this section, we consider case 1 in Eq.\,(\ref{eq:clasess}), and study the statistics of the maximum out of $N$ variables. Consider $N$ i.i.d random variables, denoted by ${x_i}$, taken from the parent PDF $p(x,t)$. To simplify, we assume the support $S$ is given by $0<x<\infty$. 
Denote the complementary cumulative distribution $s(x,t)$ by \footnote{Note that in some models the dynamics is confined to $x\in(0,x_{\max})$; consequently, $p(x>x_{\max},t)=0$, and the upper limit of the integral becomes $x_{\max}$. }
\begin{equation}\label{eq:sprob}
s(x,t)=\int_x^\infty p(u,t)\,du .
\end{equation} 
We define the sample maximum as
\begin{equation}
X_{max}=\max\{x_1,x_2,\ldots,x_N \}\, ,
\end{equation}
and study its distribution. Independence implies that the cumulative distribution function of the maximum is
\begin{equation}
Q_N^{\max}(m,t)\equiv \Pr\!\big(X_{max}<m\big)
=\big[1-s(m,t)\big]^N .
\label{eq:QN_def}
\end{equation}
We now assume the infinite-invariant density introduced in Eq.\,(\ref{eq:infden_def}). Specifically, inserting $p(x,t)\sim \mathcal{N}^{-1}t^{\alpha -1}\mathcal{I}(x)$ in Eq.\,(\ref{eq:sprob}) we integrate $\mathcal{I}(\cdot)$ to find 
\begin{equation}
s(x,t)\simeq \mathcal{N}^{-1}\, t^{\alpha-1}\,\mathcal{J}(x),
\end{equation}
where
\begin{equation}
\mathcal{J}(x)=\int_x^{\infty} \mathcal{I}(u)\,du \, . 
\label{eq:iid_scaling_tail}
\end{equation}
 Thus, at fixed $x$, $s(x,t)\to 0$ algebraically as $t\to\infty$, and the maximum statistics is controlled by the integrated infinite density $\mathcal{J}(x)$. This immediately implies that if $N$ is kept fixed and $t\rightarrow \infty$, then, trivially, $Q_N^{\max}(m,t)\to 1$, hence we should consider a different limit. On the other hand, if $N$ grows too fast with $t$ fixed, then using Eq.\,(\ref{eq:QN_def}), $Q_N^{\max}(m,t)\to 0$. The interesting regime is therefore an intermediate, nontrivial joint
limit in which the number of samples $N$ compensates for the time decay of the single-particle distribution. We therefore take the joint limit
\begin{equation}
t\to\infty,\qquad N\to\infty,\qquad 
\rho \equiv N\,t^{\alpha-1}/\mathcal{N}=\text{const},
\label{eq:rho_def_beta0}
\end{equation}
so that $N\,s(x,t)\ \longrightarrow\ \rho\,\mathcal{J}(x)$. Equivalently, one has
$s(x,t)\sim (\rho\,\mathcal{J}(x))/N$, so $s(x,t)=O(1/N)$. Inserting this expression into Eq.\,(\ref{eq:QN_def}), we obtain
\begin{equation}
\lim_{\substack{N,t \to \infty \\ \rho \text{ fixed}}}Q_N^{\max}(m,t)= \exp\!\big[-\rho\,\mathcal{J}(m)\big].
\label{eq:QN_limit_beta0}
\end{equation}
Differentiating yields the corresponding limiting density of the maximum,
\begin{equation}
q_N^{\max}(m)=\frac{\partial}{\partial m}Q_N^{\max}(m,t)
=\rho\,\mathcal{I}(m)\,\exp\!\big[-\rho\,\mathcal{J}(m)\big],
\label{eq:fM_beta0}
\end{equation}
since $\mathcal{J}'(x)=-\mathcal{I}(x)$. Finally, Eq.~\eqref{eq:QN_limit_beta0} implies the practical relation
\begin{equation}
\lim_{\substack{N,t \to \infty \\ \rho \text{ fixed}}} -\frac{1}{\rho}\ln Q_N^{\max}(m,t)\ =\ \mathcal{J}(m),
\label{eq:extract_J}
\end{equation}
where $\rho$ is fixed as given in Eq.\,(\ref{eq:rho_def_beta0}). Eq.\,(\ref{eq:extract_J}) shows how extreme-value statistics naturally relates to infinite ergodic theory via the integrated infinite
density $\mathcal{J}(x)$.

We emphasize that Eqs.\,(\ref{eq:QN_limit_beta0} - \ref{eq:extract_J}) are nontrivial. We note that Eq.\,(\ref{eq:QN_limit_beta0}) shows that part the information contained in the parent distribution $p(x,t)$ is irrelevant for the extreme statistics. In particular, in Eq.,(\ref{eq:infden_def}) we keep $x$ fixed; as shown below, this implies that the infinite invariant density $\mathcal{I}(x)$ encodes rare events. By contrast, $x$ scaling with $t$ yields typical events, also encoded in $p(x,t)$. Hence, the EVT results in Eqs.\,(\ref{eq:QN_limit_beta0} - \ref{eq:extract_J}) magnify the contribution of rare events.



\subsection{Minimum statistics controlled by $\mathcal{I}(x)$ (case 2)}
Here we study the minimum corresponding to case 2 in Eq.\,(\ref{eq:clasess}). Let us define  $X_{min}=\min\{x_1,\ldots,x_N\}$, whose cumulative distribution is
\begin{equation}\label{eq:qmindef}
Q_N^{\min}(m,t)=Pr\left(X_{min}\leq m \right)=1-\big[1-F(m,t)\big]^N,
\qquad
F(m,t)=\int_{0}^{m} p(u,t)\,du .
\end{equation}

Assume that for fixed (small) $x$ the density admits an infinite-density scaling
\begin{equation}
p(x,t)\simeq \mathcal{N}^{-1}\,t^{\alpha-1}\,\mathcal{I}(x),
\end{equation}
so that
\begin{equation}\label{eq:Jsecmin}
F(x,t)\simeq \mathcal{N}^{-1}\,t^{\alpha-1}\,\mathcal{J}(x),
\qquad
\mathcal{J}(x)=\int_{0}^{x} \mathcal{I}(x')\,dx' .
\end{equation}
Taking the joint limit $N\to\infty$, $t\to\infty$ with $\rho$ fixed using Eq.\,(\ref{eq:rho_def_beta0}), yields
\begin{equation}\label{eq:qnmingen}
\lim_{\substack{N,t \to \infty \\ \rho \text{ fixed}}} Q_N^{\min}(m)=\;
1-\exp\!\big[-\rho\,\mathcal{J}(m)\big].
\end{equation}
Consequently, the minimum PDF is
\begin{equation}\label{eq:19}
q_N^{\min}(m)=\frac{d}{dm}Q_N^{\min}(m)
\simeq \rho\,\mathcal{I}(m)\,\exp\!\big[-\rho\,\mathcal{J}(m)\big] \, .
\end{equation}
Comparing with the PDF of the  maximum in Eq.\,(\ref{eq:fM_beta0}) we see the analogy, here we replace the integrated tail
$\int_x^\infty \mathcal{I}(x')\,dx'$ with the small-$x$ integral $\mathcal{J}(x)=\int_0^x \mathcal{I}(x')\,dx'$.

Our central message is that, once the long-time single-particle statistics are described by an infinite invariant density,
the corresponding extreme-value statistics in the joint limit of long time and large sample size can be obtained in a closed form either for the maximum in Eq.~(\ref{eq:QN_limit_beta0}) or for the minimum in Eq.\,(\ref{eq:qnmingen}). 

\section{Model illustrations}
\label{sec:scope_models}
While the main Eqs.\,(\ref{eq:fM_beta0}) and (\ref{eq:19}) are valid in the limit of $t,N \rightarrow \infty$, as explained above, one remaining question is what would happen for finite $t,N$ while both parameters are large. Below we study three models using analytical arguments and finite-time simulations to verify the applicability of the theory.


\subsection{Overdamped Langevin particle in an asymptotically flat potential}\label{sec:Boltzmann}
\subsubsection{The infinite invariant density}
We consider the overdamped motion of a Brownian particle on the half-line $x>0$ in an external potential $V(x)$, in contact with a heat bath at temperature $T$ \cite{aghion2019non}. The dynamics is governed by the Langevin equation
\begin{equation}
\label{eq:SDE}
\frac{dx}{dt} \;=\; -\frac{1}{\gamma}\,V'(x) \;+\; \sqrt{2D}\,\eta(t),
\qquad
\langle \eta(t)\eta(t')\rangle=\delta(t-t'),
\end{equation}
where $\gamma>0$ is the friction coefficient, and $\eta(t)$ is standard Gaussian white noise with $\langle \eta(t)\rangle=0$ and $\langle \eta(t)\eta(t')\rangle=\delta(t-t')$. Here $V(x)$ is the external potential and $V'(x)=dV/dx$. The bath temperature $T$ fixes the diffusion constant $D$ through the Einstein relation $D = k_B T/\gamma$, where $k_B$ is Boltzmann’s constant. We assume a localized initial condition at $x_0>0$ (or any initial density with sufficiently fast-decaying tails), the long-time results below are insensitive to this choice \cite{aghion2020infinite}. Throughout this section we adopt the one-sided geometry with a reflecting boundary at the origin (either as an explicit wall or via a divergence of $V(x)$ as $x\rightarrow 0$).

The probability density $p(x,t)$ obeys the corresponding Fokker--Planck equation
\begin{equation}
\label{eq:FP}
\partial_t p(x,t)
= -\partial_x \big[ v(x)\,p(x,t)\big] + D\,\partial_x^2 p(x,t),
\qquad
v(x) \equiv -\frac{1}{\gamma} V'(x),
\end{equation}
supplemented by the no-flux boundary condition at $x=0$,
\begin{equation}
\label{eq:noflux}
J(0,t)=0,
\qquad
J(x,t)=v(x)\,p(x,t)-D\,\partial_x p(x,t).
\end{equation}
We are interested in asymptotically flat potentials, for which
\begin{equation}
\label{eq:asympflat}
V'(x)\to 0
\quad\text{and}\quad
V(x)\to 0
\qquad (x\to\infty).
\end{equation}
In this case the Boltzmann weight $\exp\left[{-V(x)/k_B T}\right]$ is non-normalizable on $[0,\infty)$, so a conventional equilibrium density does not exist \cite{farago2021thermodynamics}. Note that, although the Boltzmann weight is non-normalizable, it is still a time-independent solution of of the Fokker-Planck Eq.\,(\ref{eq:FP}). For long times the parent distribution $p(x,t)$ admits a uniform approximation that consistently matches the near-field (potential-dominated) region with the large-$x$ region \cite{aghion2020infinite},
\begin{equation}
\label{eq:uniform}
p(x,t)\;\simeq\;\frac{1}{\sqrt{\pi D t}}\,
\exp\!\Bigg[-\,\frac{V(x)}{k_B T}\;-\;\frac{x^2}{4Dt}\Bigg],
\qquad x>0,
\end{equation}
where the diffusive Gaussian factor $\exp{\left( -x^2/4Dt\right)}$ controls the $x\sim \sqrt{Dt}$ scales. As we show below, for the calculation of the minimum, this Gaussian factor is an irrelevant part of the parent distribution $p(x,t)$ that can be ignored.

Defining the time-dependent scaling factor
\begin{equation}
\label{eq:Zt}
Z_t \;=\; \sqrt{\pi D t}\, ,
\end{equation}
and using Eq.\,({\ref{eq:infden_def}}), we see that $\mathcal{N} = \sqrt{\pi D}$ and $\alpha=1/2$. From Eq.\,(\ref{eq:uniform}), the Boltzmann infinite-density limit is therefore 
\begin{equation}
\label{eq:infinite-density_bol}
\mathcal{I}(x)\equiv\lim_{t\to\infty} Z_t\,p(x,t)
\;=\;
\exp\!\Big[-\,\frac{V(x)}{k_B T}\Big],
\end{equation}
i.e.\ the rescaled density $Z_t p(x,t)$ converges to the non-normalizable Boltzmann weight. 
Importantly, $\alpha=1/2$ is the return exponent. Roughly speaking, after escaping the potential well and reaching the region where $V\sim 0$, the distribution of return times follows that of a free particle. In this regime, well known Lévy–Smirnov behavior is expected, so that the survival probability decays as $ \propto t^{-1/2}$ and consequently $\alpha = 1/2$, as anticipated \cite{schrodinger1915theorie}.

\paragraph*{The Lennard--Jones potential}

\begin{figure}[h]
        \centering
\includegraphics[width=0.5\linewidth]{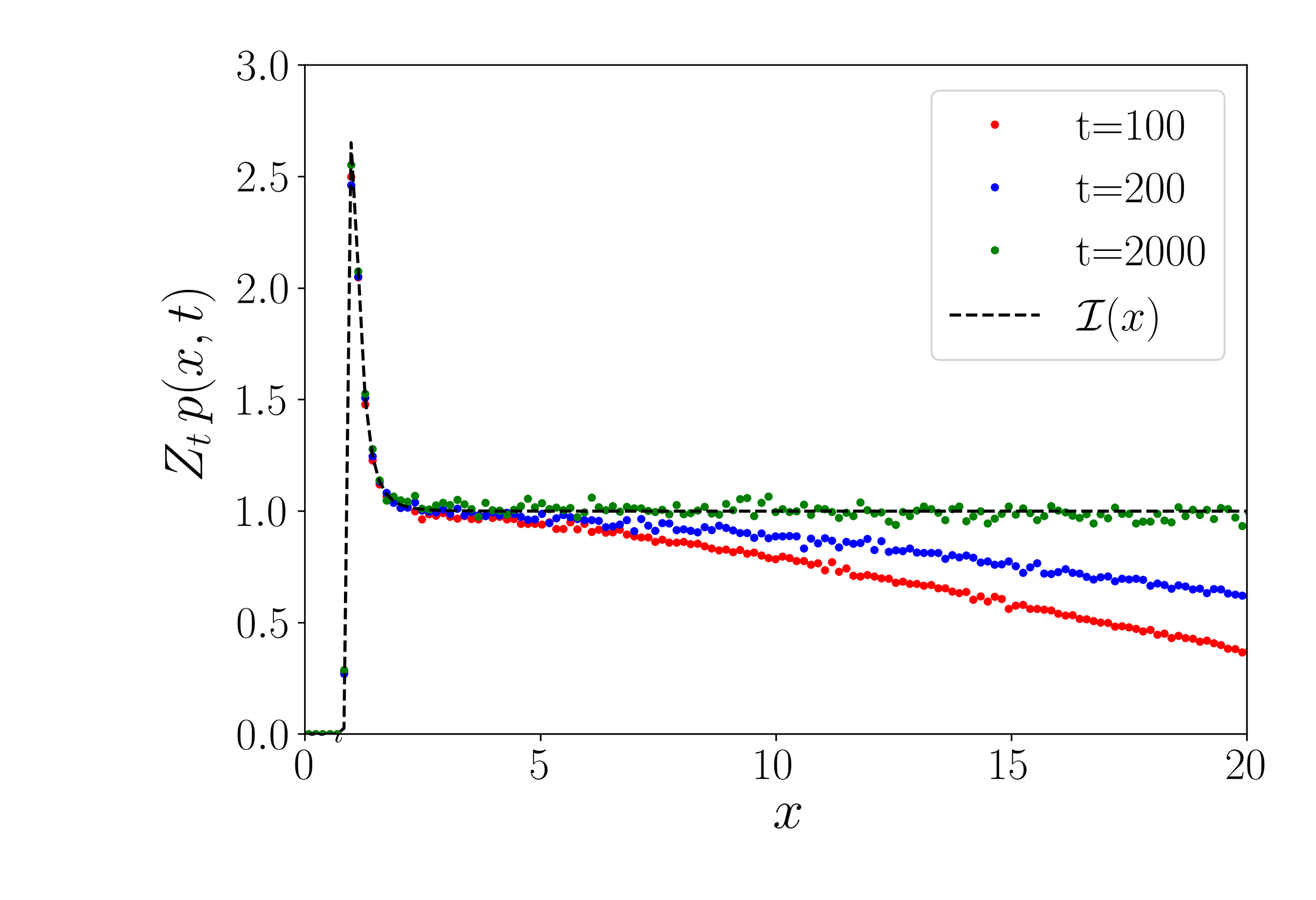}
    \caption{ 
Convergence of the rescaled single-particle PDF to an infinite invariant density for a particle in an asymptotically flat Lennard-Jones potential in Eq.\,(\ref{eq:LJ_tail}) with the parameters in Eq.\,(\ref{eq:parameterspot}) \cite{aghion2020infinite}.
As $t$ increases, the distribution converges to the infinite-density scaling form $\mathcal{I}(x)$ given in Eq.\,(\ref{eq:boltexample}), which is the Boltzmann state in a non normalized form. 
}
\label{fig:Boltzmann_infinite_density}
\end{figure}

To provide a concrete illustration, we consider a one-sided asymptotically flat potential
of Lennard--Jones type,
\begin{equation}
V(x)=\frac{A}{x^{n}}-\frac{B}{x^{\mu}},
\qquad A,B>0,\qquad n>\mu>1,\qquad x>0, 
\label{eq:LJ_tail}
\end{equation}
which is strongly repulsive near the origin and satisfies $V(x)\to 0$ as $x\to\infty$. The infinite density obtained in Eq.~(\ref{eq:infinite-density_bol}) reads explicitly
\begin{equation}\label{eq:boltexample}
\mathcal{I}(x)=\exp\!\left[-V(x)/(k_B T)\right]
=\exp\!\left[-\frac{1}{k_BT}\left(\frac{A}{x^{n}}+\frac{B}{x^{\mu}}\right)\right].
\end{equation}
Hence clearly in the limit $x\rightarrow \infty$, $\mathcal{I}(x)\rightarrow 1$, implying $\beta=0$ for case 2 in Eq.\,(\ref{eq:clasess}).
In the numerical examples below we use
\begin{equation}\label{eq:parameterspot}
A=1,\qquad B=2,\qquad n=12,\qquad \mu=6\, . 
\end{equation}
The convergence of the rescaled density $Z_t p(x,t)$ toward this non-normalizable
Boltzmann weight is demonstrated in Fig.~\ref{fig:Boltzmann_infinite_density}. We see a peak for $\mathcal{I}(x)$ close to the minimum of the potential.
For further details see \cite{aghion2019non,farago2021thermodynamics}.

 \subsubsection{Extreme value statistics for the minimum}
\label{sec:Boltzmann_minimum}
As mentioned, $\mathcal{I}(x)\to 1$ as $x\to\infty$, and hence the extreme value theory of case 2 applies, therefore, the appropriate extreme to consider is the minimum position of $N$ independent particles. We apply the general minimum extreme-value theory derived in Eqs.\,(\ref{eq:qnmingen}) and (\ref{eq:19}) to the above Langevin dynamics in the present
asymptotically flat potential. Although most of the particles drift to large distances
as $t$ grows, the minimum is controlled by the rare particles that 
remain in the
inner, potential-dominated region $x=O(1)$ where $x\geq 0$.  Following Eq.\,(\ref{eq:infinite-density_bol}) the single-particle small-$x$ CDF satisfies
\begin{equation}
F(x,t)\equiv \int_0^x p(u,t)\,du
\simeq \frac{1}{Z_t}\,\mathcal{J}(x)\,,
\end{equation}
where $\mathcal{J}(\cdot)$ is defined using Eqs.\,(\ref{eq:iid_scaling_tail}) and Eq.\,(\ref{eq:infinite-density_bol}).
 Let
 \begin{equation}
X_{\min}=\min\{x_1,\ldots,x_N\}\, \,
\end{equation}
and $Q_N^{min}(m,t)$ is defined in Eq.\,(\ref{eq:qmindef}). In the joint limit $N,t\to\infty$ with 
\begin{equation}\label{eq:boltzmannqn}
\rho \equiv \frac{N}{Z_t}=\frac{N}{\sqrt{\pi D t}}
\end{equation}
held fixed, the distribution of the minimum converges to the nontrivial
limit,
\begin{equation}\label{eq:boltzmannQN}
Q_N^{\min}(m,t)\;\longrightarrow\;1-\exp\!\big[-\rho\,\int_0^m du\exp\!\Big[-\,\frac{V(u)}{k_B T}\Big]\big]\,,
\end{equation}
and
\begin{equation}
q_N^{min}(m,t)\equiv \partial_m Q_N^{\min}(m,t)\;\longrightarrow\;
\rho\,\exp\!\Big[-\,\frac{V(m)}{k_B T}\Big]\,\exp\!\big[-\rho\,\int_0^m du\exp\!\Big[-\,\frac{V(u)}{k_B T}\Big]\big].
\end{equation}
Thus, for finite $\rho$ the statistics of the minimum are governed directly by the
infinite density $\mathcal{I}(x)$.

\begin{figure}[h]
    \centering
    \begin{minipage}[t]{0.49\linewidth}
        \centering
    \includegraphics[width=\linewidth]{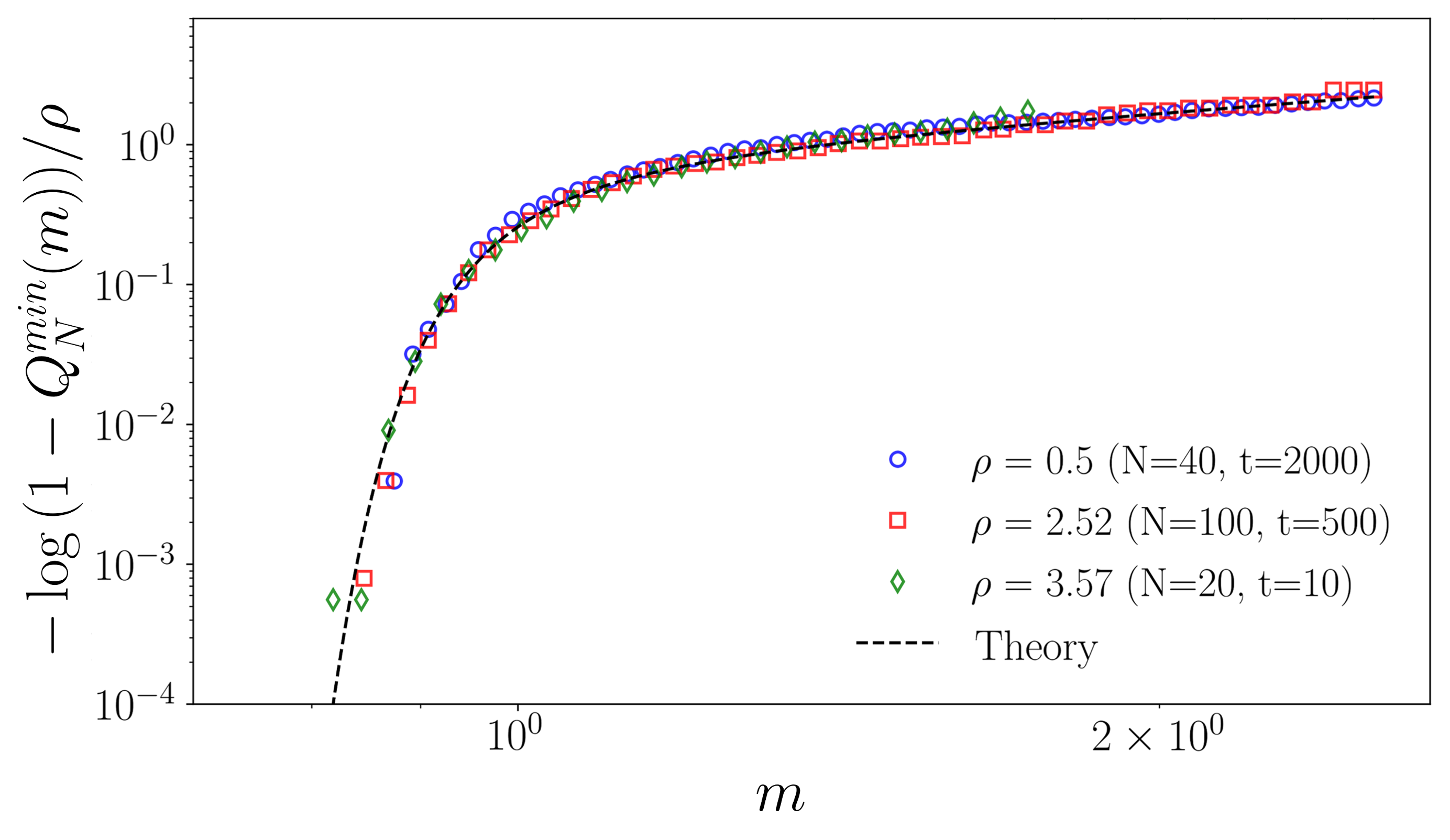}
        \caption{Numerical test of the prediction in Eq.\,(\ref{eq:boltzmannQN}) for the minimum CDF $Q_N^{min}(m,t)$ for overdamped diffusion in an asymptotically flat potential of Lennard-Jones type in Eq.\,(\ref{eq:LJ_tail}). Shown is $-\ln[1-Q_N^{min}(m,t)]/\rho$ versus $m$ for several effective densities $\rho=N/\sqrt{\pi Dt}$ obtained from Langevin simulations. In the limit $N,t\to\infty$ at fixed $\rho$, all curves collapse onto the integrated infinite density $\mathcal{J}(\cdot)$ (black dashed line), where $\mathcal{I}(\cdot)$ is the non-normalizable Boltzmann weight given in Eq.\,(\ref{eq:boltexample}). Here we used $D=1$ and $k_BT=1$,  with the parameters in Eq.\,(\ref{eq:parameterspot}).}
\label{fig:boltzmanncollapse}
    \end{minipage}
    \hfill
    \begin{minipage}[t]{0.49\linewidth}
        \centering
    \includegraphics[width=\linewidth]{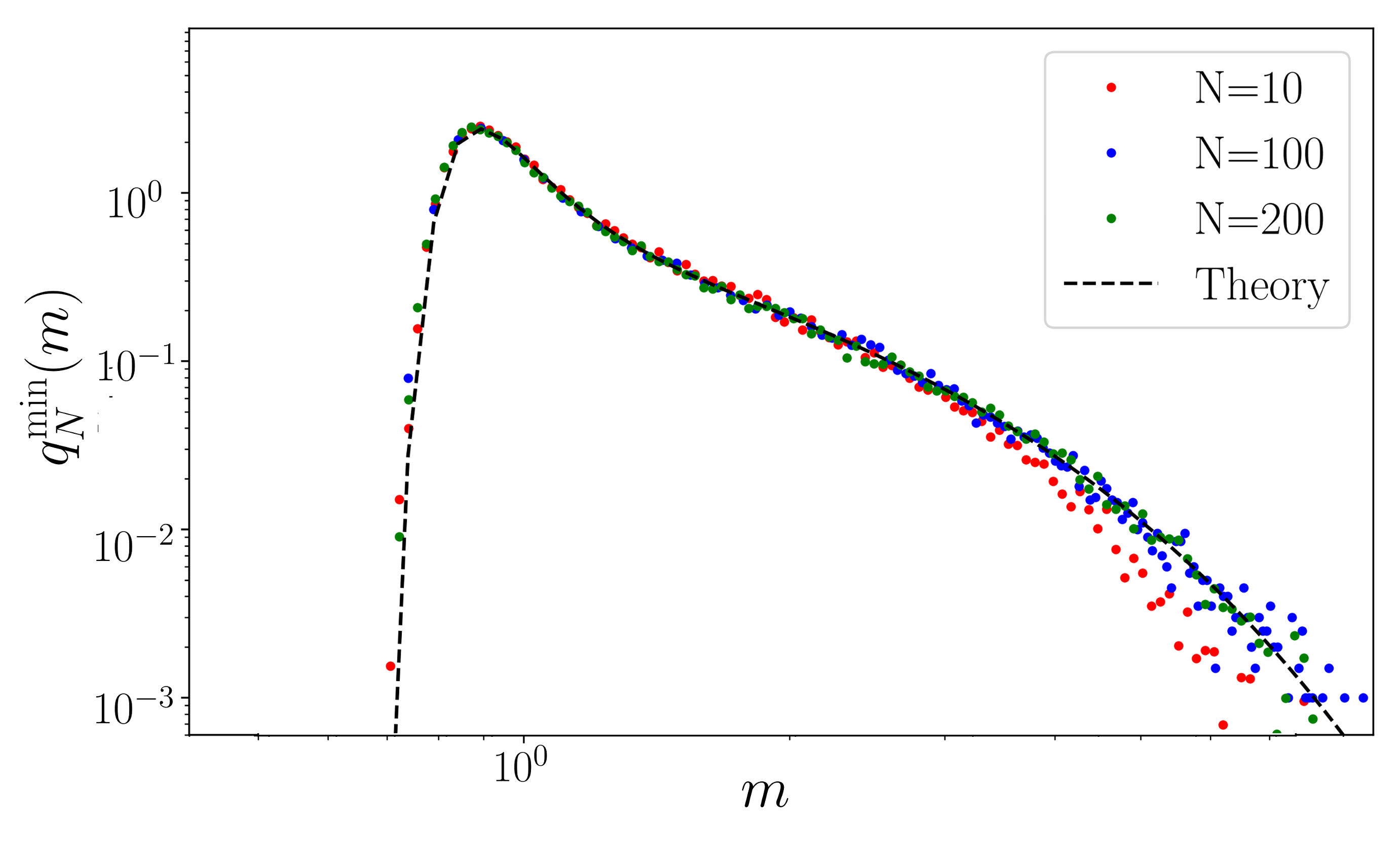}
    \caption{Probability density $q_N^{min}(m,t)$ of the minimum position for $N$ random variables, shown for several $N$ (with $t$ chosen such that $\rho=2$ is fixed). The symbols, represent how data obtained from Langevin simulations, approach the limiting extreme-value form (black dashed line). The model uses an Lennard-Jones potential in Eq.\,(\ref{eq:LJ_tail}). We see that observing very small $m$ is very unlikely due to diverging character of the potential field at $x\rightarrow 0$. We use the same parameters as in Fig. \ref{fig:boltzmanncollapse}. 
}\label{fig:boltzmannqn}
    \end{minipage}
\end{figure}

This prediction for the minimum is tested numerically in two ways. First, we show
the collapse of $-\ln[1-Q_N(m,t)]/\rho$, following Eq.\,(\ref{eq:boltzmannQN}), plotted versus $x$ for several values of $\rho$,
confirming that it converges to the integrated infinite density $\mathcal{J}(m)$ (see
Fig.~\ref{fig:boltzmanncollapse}). Second, we present the corresponding minimum PDF for a
representative value, $\rho=2$, demonstrating agreement with Eq.\,(\ref{eq:19}) (see Fig.~\ref{fig:boltzmannqn}).

\begin{figure}[h]
        \centering
\includegraphics[width=0.5\linewidth]{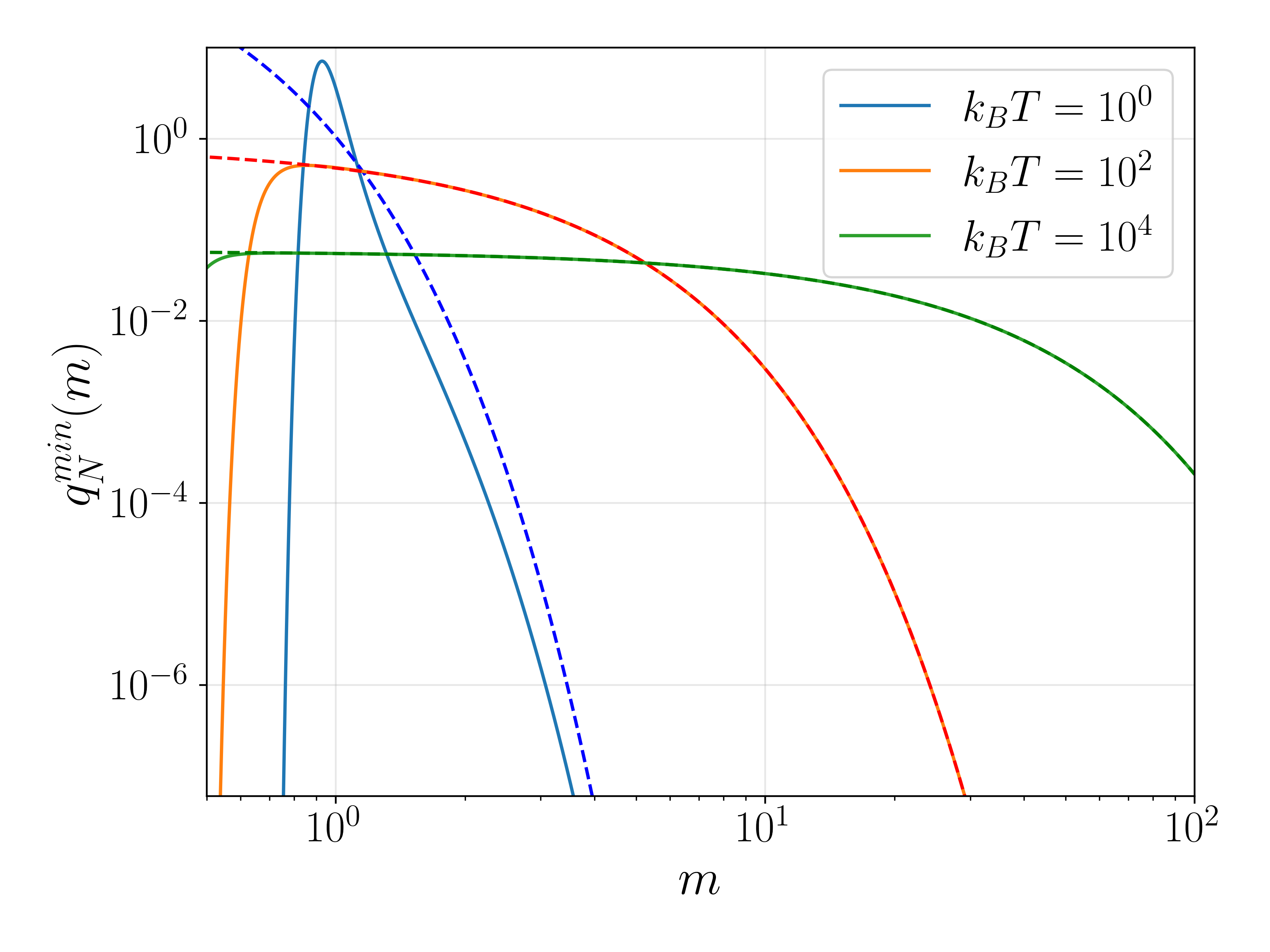}
    \caption{Comparison of the limiting distribution $q_N^{min}(m,t)\to \rho\,\mathcal{I}(m)\exp[-\rho \mathcal{J}(m)]$ (solid lines) for different temperatures for $N=100,t=100$ for the overdamped Langevin model. The figure shows that at lower temperatures, the PDF of the minimum position has a peak close to the minimum point of the potential. The distribution becomes progressively flatter as $T$ increases, approaching the free-diffusion form in the high-temperature limit. The dashed lines indicate the high-temperature limit approximation in Eq.\,(\ref{eq:hightwithc}). Agreement with this approximation improves as $T$ increases, while deviations persist at small $m$ due to the divergent repulsive core of $V(x)$. }
\label{fig:Boltzmann_highT}
\end{figure}

Since temperature enters the minimum statistics through the infinite density $\mathcal{I}(x)=\exp[-V(x)/(k_B T)]$. At fixed $\rho$, lowering $T$ makes the potential-induced feature in $q_N ^{min}(m,t)$ (the peak related to the minimum of the potential) more pronounced. As $T$ increases, thermal fluctuations facilitate escape from the well region, the PDF of the minimum becomes flatter, and this feature is progressively smoothed out. This trend is shown in Fig.\,\ref{fig:Boltzmann_highT}, where the bump in $q_N^{min}(m,t)$ is strongest at low temperature and becomes flatter as $T$ increases. In the high-temperature limit, $\mathcal{I}(x)\to 1$ and therefore $\mathcal{J}(x)
\sim x - c(T)$. For the Lennard–Jones potential, one can show that $c(T)=-\frac{1}{12}\,\Gamma\!\left(-\frac{1}{12}\right)
\left(k_B T\right)^{-1/12}
-\frac{1}{6}\,\Gamma\!\left(\frac{5}{12}\right)\,(k_B T)^{-7/12}
$ for the parameters in Eq.\,(\ref{eq:parameterspot}). Note that due to the diverging potential at $x\rightarrow 0$, this approximation will only be valid for larger $m$. Substituting this into the extreme-value theory yields the
high-temperature approximation $Q_N^{min}(m,t)\simeq 1-\exp\!\big[-\rho\,(m-c(T))\big]$, therefore
\begin{equation}\label{eq:hightwithc}
    q_N^{min}(m,t)\simeq \frac{N}{\sqrt{\pi D \,t}}\,\exp\!\big[-\frac{N}{\sqrt{\pi D \,t}}\,(m-c(T))\big] \, .
\end{equation}
As shown in Fig.~\ref{fig:Boltzmann_highT}, this approximation improves as $T$,
increases but performs worse close to $m\rightarrow 0$, where the Lennard–Jones potential diverges. Specifically, $c(T)\to 0$ as $T\to\infty$, so that the limiting form becomes simply $Q_N^{\min}(m,t)\xrightarrow[T\to\infty]{} 1-\exp(-N m/\sqrt{\pi D \,t})$.

\subsection{Weakly chaotic maps: intermittency and infinite invariant density}
\label{sec:weak_maps}
\subsubsection{The infinite invariant density}
In this subsection, we study weakly chaotic intermittent maps and analyze the extreme-value statistics of the maximum of $N$ observations of the relevant observable $x$. Weakly chaotic (intermittent) maps provide a canonical deterministic setting in which the usual ergodic picture breaks down due to the presence of a marginally unstable fixed point \cite{pomeau1980intermittent,thaler1983transformations,thaler2000asymptotics,meyer2018anomalous}. A widely studied example is the Pomeau–Manneville (PM) map, introduced to describe intermittency in the transition to turbulence.
A prototypical scenario is an interval map $M:[0,1]\to[0,1]$ iterated as
\begin{equation}
x_{t+1}=M(x_t),
\end{equation}
with discrete time $t=0,1,2,\ldots$, where trajectories spend long laminar episodes near the marginal point, interrupted by fast
excursions across the interval \cite{gaspard1988sporadicity}. This mechanism yields vanishing (standard) Lyapunov exponent and
subexponential separation of trajectories, and it places the dynamics naturally within the
framework of infinite ergodic theory and its statistical consequences
(e.g.\ Aaronson--Darling--Kac type limits, generalized Pesin identities, and related constructions) \cite{korabel2013numerical,korabel2009pesin,brevitt2025does,pomeau1980intermittent,thaler1983transformations,thaler2000asymptotics,aaronson1997introduction,brevitt2025singularity,sandev2026anomalous,endo2025noise}.

A large class of such maps is characterized by the local expansion near the marginal fixed point
(at $x=0$), such as the Pomeau–Manneville (PM) maps of the form 
\begin{equation}\label{eq:pmmap}
M_{\rm PM}(x)\sim x+a x^{z}\ ,
\end{equation}
for $x\rightarrow 0 $ with $a>0$ and $z>2$. Let $p(x,t)$ be the ensemble density after $t$ iterations, which can be obtained from the Frobenius--Perron operator using a smooth initial density $p(x,0)$, see appendix \ref{appendix:FP}. In the weakly chaotic regime ($z>2$) the density does not
converge to a normalizable invariant measure. Instead, there exists an exponent
\begin{equation}\label{eq:alphadef}
\alpha=\frac{1}{z-1}\in(0,1),
\end{equation}
so the infinite-density defined in Eq.\,(\ref{eq:infden_def}) with $\mathcal{N}=1$ is given by
\begin{equation}
\mathcal{I}(x)=\lim_{t\to\infty} t^{1-\alpha} p(x,t).
\label{eq:IID_maps}
\end{equation}
The parameter $\alpha$ controls the statistics of laminar episodes near the marginal fixed point at $x=0$.
To quantify their duration, we introduce the first-exit time from a small neighborhood $(0,b)$, with fixed $b\in(0,1)$,
\begin{equation}
\tau_b=\inf\{t\ge 0:\ x_t\ge b\},
\qquad
\Pr(\tau_b>t)\sim t^{-\alpha}\quad (t\to\infty).
\label{eq:survival}
\end{equation}
A continuous-time approximation of Eq.~(\ref{eq:IID_maps}) yields the power-law survival probability in Eq.~\eqref{eq:survival}, and $\alpha$ is the return/persistence exponent \cite{zumofen1993scale,geisel1984anomalous}.

In particular, the infinite density in Eq.\,(\ref{eq:IID_maps}) singular structure is algebraic 
\begin{equation}\label{eq:IID_maps2}
\mathcal{I}(x)\sim B\,x^{-1/\alpha},
\qquad \,.
\end{equation}
 for small-$x$ region, hence this case corresponds to class 1 in our scheme Eq.\,(\ref{eq:clasess}) with $\beta=1/\alpha$. This divergence reflects the intermittent dynamics: trajectories linger for long times in
the vicinity of the marginally unstable fixed point at $x=0$, so probability weight
continuously accumulates near the origin. The accumulation is so strong that
$\mathcal{I}(x)$ is not integrable at $x=0$, i.e.\ it defines an infinite (non-normalizable)
invariant density. Moreover, Thaler's theorem shows that the scaling holds far beyond specific examples: it applies to a broad class of
interval maps with a single marginal fixed point at the origin \cite{thaler2000asymptotics}.

\paragraph*{Pomeau--Manneville dynamics}
For the PM maps in Eq.\,(\ref{eq:pmmap}), one does not have a closed-form expression for $\mathcal{I}(x)$ on the full interval $(0,1)$, but the small-x region described by the algebraic divergence in Eq.\,(\ref{eq:IID_maps2}) is specified by
\cite{korabel2009pesin,korabel2013numerical}
\begin{equation}\label{eq:bcoeff}
B = \frac{\sin(\pi\alpha)}{\pi\alpha}\left(\frac{a}{\alpha}\right)^{\alpha-1} \, .\end{equation} 
Here $B$ is fixed by our operational definition in Eq.\,(\ref{eq:IID_maps}).

\begin{figure}[h]
    \centering    \includegraphics[width=0.6\linewidth]{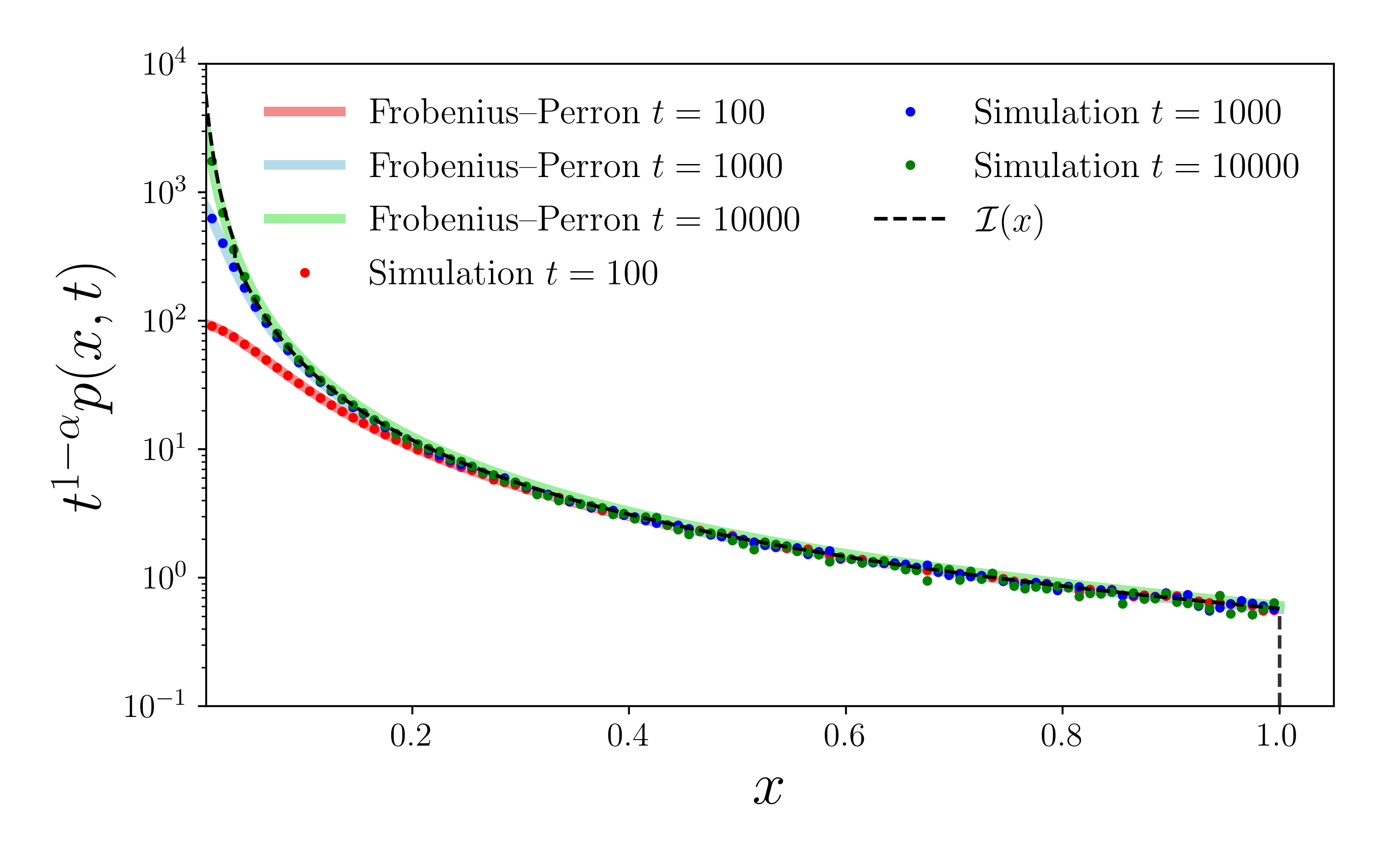}
        \caption{
Rescaled density $t^{1-\alpha}p(x,t)$ of the Thaler map with $z=3$, evaluated at $t=10^2,10^3,10^4$ compared to the infinite invariant density $\mathcal{I}(x)$ (black dashed curve) given in Eq.\,(\ref{eq:thaler_I}). Solid curves are obtained from a numerical iteration of the Frobenius–Perron operator (see Appendix \ref{appendix:FP} for details). Dots show direct Monte Carlo simulations. See \cite{korabel2013numerical} for further details.
}
\label{fig:map_infinite_density}
    \end{figure}

An analytically tractable example of the PM dynamics is the Thaler map
\begin{equation}\label{eq:thalermap}
M_{\rm Th}(x)=x\left[1+\left(\frac{x}{1+x}\right)^{z-2}-x^{z-2}\right]^{-1/(z-2)} \pmod 1,
\end{equation}
 where $\alpha$ is given in Eq.\,(\ref{eq:alphadef}) and $a=1$. Eq.\,(\ref{eq:thalermap}) has two inverse branches and the same marginal behavior Eq.~\eqref{eq:pmmap} at $x\to 0$.
For this map for, an explicit infinite invariant density is known \cite{thaler2000asymptotics,korabel2013numerical}:
\begin{equation}
\mathcal{I}(x)=B\Big[x^{-1/\alpha}+(1+x)^{-1/\alpha}\Big],
\label{eq:thaler_I}
\end{equation}
where $B$ is given by Eq.\,(\ref{eq:bcoeff}). To test the theory, one may start from a smooth initial density (e.g.\ uniform on
$(0,1)$); the long-time rescaled density $t^{1-\alpha}p(x,t)$ then converges to $\mathcal{I}(x)$.
This behavior is illustrated in Fig.~\ref{fig:map_infinite_density}, where the rescaled density
$t^{1-\alpha}p(x,t)$ for the Thaler map approaches the infinite invariant
density $\mathcal{I}(x)$ in Eq.\,(\ref{eq:thaler_I}).

For finite long $t$, the single-particle density $p(x,t)$ contains both a regular laminar-core sector ($x \lesssim x_c(t)$), and the infinite-density sector ($x > x_c(t)$). The boundary $x_c(t)\sim \alpha^{\alpha}t^{-\alpha}$ approaches zero when $t\rightarrow \infty$. 
For the Thaler map, for $x<x_c(t)$, the solution is given by \cite{akimoto2013aging}
\begin{equation}
    p(x,t)\simeq \frac{1}{x_c(t)}\,\Phi\!\left(x/x_c(t)\right) \qquad \Phi(u)=\frac{\sin(\pi\alpha)}{\pi\alpha}\,\frac{1}{1+u^{1/\alpha}} \, 
    \label{eq:aki}
\end{equation}
instead by the infinite invariant density in Eq.\,(\ref{eq:thaler_I}).
Eq. (\ref{eq:aki}) regularizes the small-$x$ core of $p(x,t)$ while matching the infinite-density form for
$x \gg x_c(t)$. Note for the PM map, the same interpretation applies, although a closed-form
expression for $\mathcal{I}(x)$ on the full interval is not available, yet one can evaluate this function using the Frobenius–Perron operator (see appendix \ref{appendix:FP}).
For the fixed-$\rho$ extreme-value theory considered below, however, the maximum is controlled
by rare outer excursions, whereas the laminar core corresponds to common events. Since
$x_c(t)\to 0$ as $t\to\infty$, one has $\Phi(x/x_c(t))\to 1$ at any fixed $x>0$. As a consequence, extreme-value statistics of the maximum is solely determined by the infinite invariant density
$\mathcal{I}(x)$, while the function $\Phi(u)$ is irrelevant.

\subsubsection{Statistics of 
the maximum $X_{\max}$}
\begin{figure}[h]
    \centering
    \begin{minipage}[t]{0.49\linewidth}
        \centering
    \includegraphics[width=\linewidth]{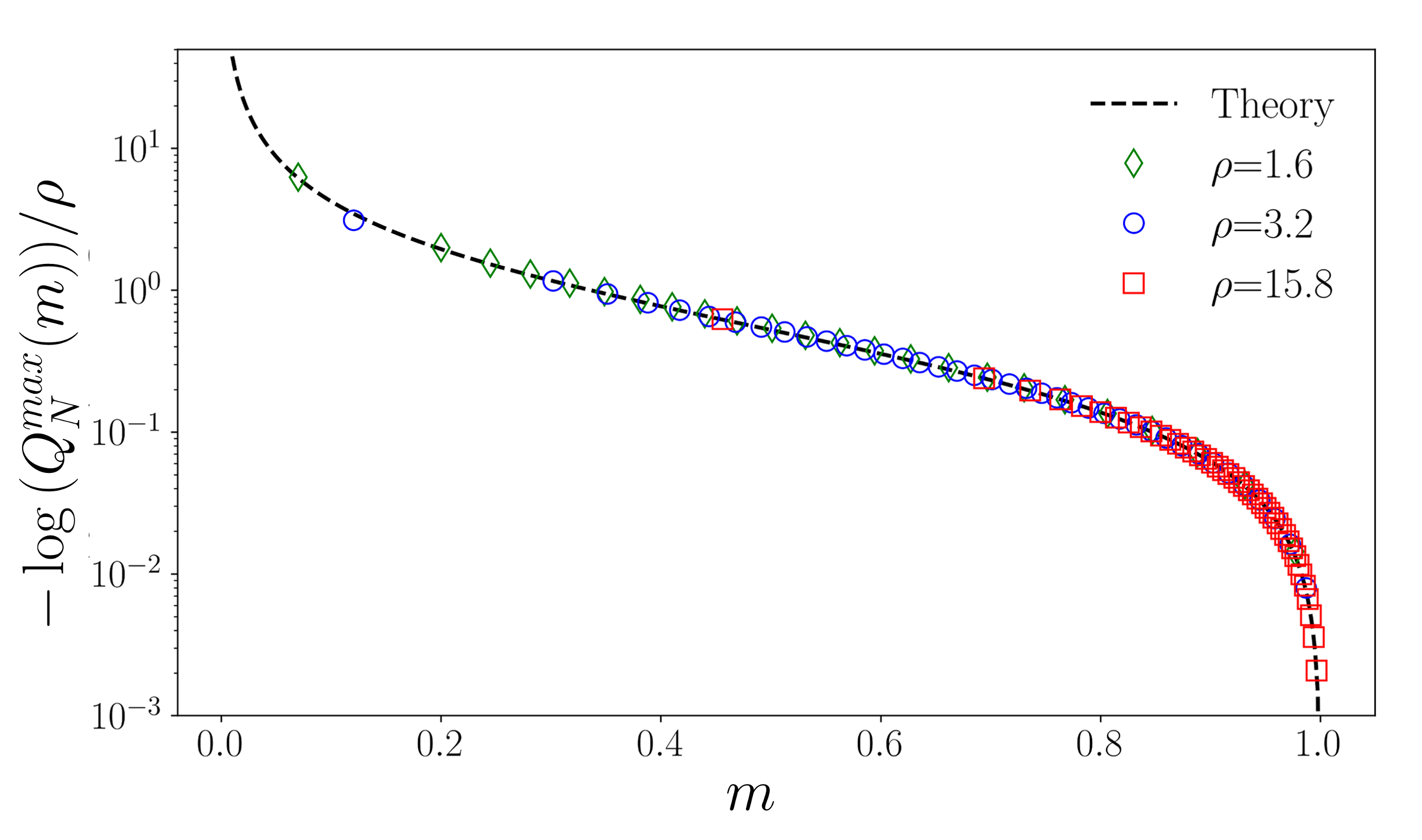}
        \caption{Data collapse of the maximum $x$ for the Thaler map with $z=3$. Symbols are Monte Carlo data for $t=10^3$ iterations. In each realization, $N$ initial conditions were drawn independently and uniformly on $[0,1]$, iterated under the map for $t$ steps, and the maximum $X_{\max}$ was recorded.
        The curves collapse onto the integrated infinite invariant density $\mathcal{J}(x)=\int_x^1 \mathcal{I}(u)\,du$ (dashed line), where $\mathcal{I}(\cdot)$ is given in Eq.\,(\ref{eq:thaler_I}).}
\label{fig:mapcollapse}
    \end{minipage}
    \hfill
    \begin{minipage}[t]{0.49\linewidth}
        \centering
    \includegraphics[width=\linewidth]{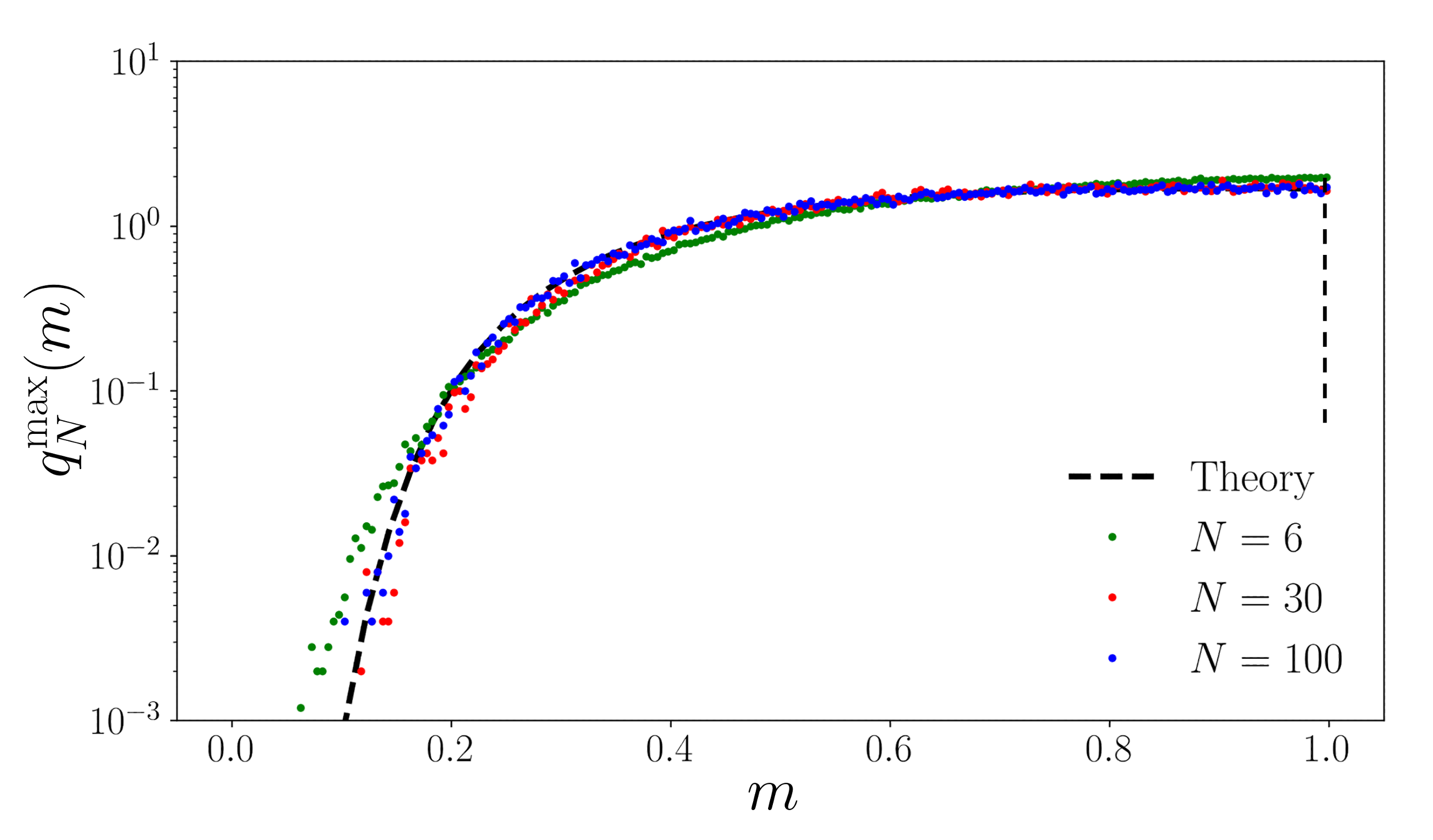}
    \caption{Convergence of the maximum PDF at fixed $\rho$ for the Thaler map with $z=3$. Shown is the probability density $q_N^{max}(m,t)$ for several $(N,t)$ satisfying $\rho=N t^{\alpha-1}=3$. As $N,t\to\infty$ at fixed $\rho$, the numerical curves converge to the limiting form $q_N^{max}(m)$ given in Eq.\,(\ref{eq:qformap}) (dashed line). For each realization, $N$ initial conditions were drawn independently from a uniform distribution on $[0,1]$, iterated under the map for $t$ steps, and the maximum was recorded; the procedure was repeated $M=5\times 10^{5}$ times to build the histogram.} \label{fig:mapqn}
    \end{minipage}
\end{figure}

As mentioned, the infinite invariant density $\mathcal{I}(x)$ diverges as $x\to 0$, which makes it non-integrable and places the map in the non-normalizable class 1 mentioned in Eq.\,(\ref{eq:clasess}). The extreme statistics we consider here concerns rare large excursions far from the unstable fixed point of the observable, and is therefore governed by the maximum. We consider $N$ independent realizations of the map, all evolved for $t$ iterations, and denote by
$x_i(t)$ the position of the $i$th trajectory at time $t$. We study the sample maximum
\begin{equation}
X_{\max}=\max\{x_1,\ldots,x_N\}\, .
\end{equation}
Its cumulative distribution is defined in Eq.\,(\ref{eq:QN_def}) with
\begin{equation}
s(x,t)=\int_x^{1} p(u,t)\,du ,
\end{equation}
where the upper limit $1$ reflects the unit-interval support of the map $x\in [0,1]$, namely $X_{max}\leq 1$. Using the infinite-density established in Eq.\,(\ref{eq:IID_maps}), we obtain
\begin{equation}
s(x,t)\sim t^{\alpha-1}\,\mathcal{J}(x),
\qquad
\mathcal{J}(x)\equiv \int_x^{1}\mathcal{I}(u)\,du .
\end{equation}
Taking the joint limit $N\to\infty$, $t\to\infty$ with the ratio
\begin{equation}
\rho \equiv N\,t^{\alpha-1}
\qquad
\Big(\alpha=\frac{1}{z-1}\Big)
\end{equation}
held fixed, the general extreme-value result of Sec.~\ref{subsec:ev} yields , following Eqs.\,(\ref{eq:QN_limit_beta0}) and (\ref{eq:fM_beta0})
\begin{equation}\label{eq:qformap}
    q_N^{max}(m)=\rho\, B\Big[m^{1-z}+(1+m)^{1-z}\Big]\exp\left(-\rho \int_m^1 du \,B\Big[u^{1-z}+(1+u)^{1-z}\Big] \right) \, .
\end{equation}
The collapse for different values of $\rho$
of $-\ln Q_N^{max}(m,t)/\rho \to \mathcal{J}(x)$ in Eq.\,(\ref{eq:QN_limit_beta0}) is tested in Fig.~\ref{fig:mapcollapse},
while Fig.~\ref{fig:mapqn} demonstrates convergence of the maximum PDF $q_N^{max}(m,t)$
to the predicted limiting form for $\rho=3$.

\begin{figure}[h]
    \centering
    \begin{minipage}[t]{0.49\linewidth}
        \centering
    \includegraphics[width=\linewidth]{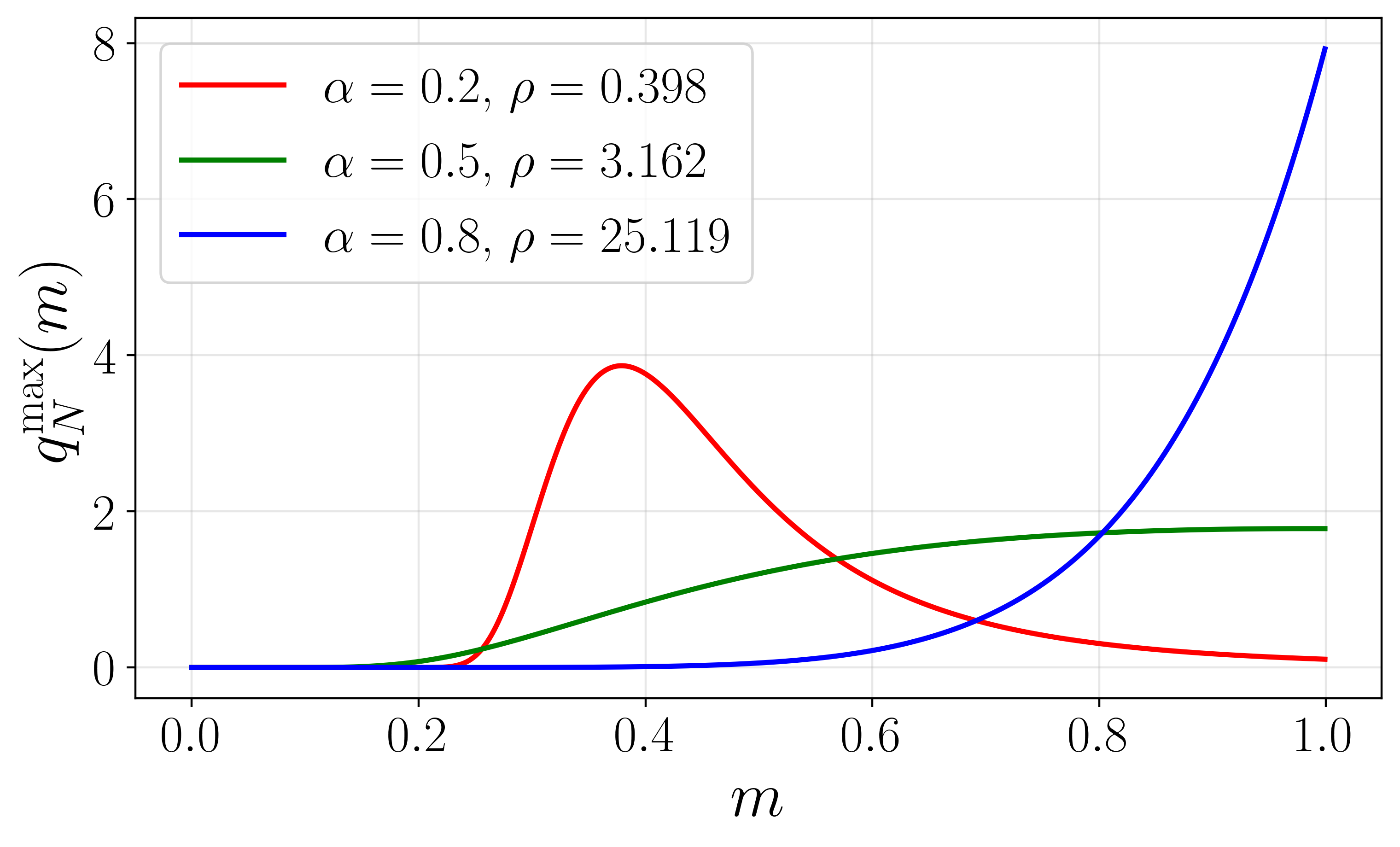}
        \caption{Plot of the solution in Eq.~(\ref{eq:qformap}) for the Thaler map, shown for $\alpha=0.2,0.5,0.8$ at fixed $N=100$ and $t=1000$. Since $\rho=N t^{\alpha-1}$ depends on $\alpha$, the three curves correspond to different values of $\rho$. With increasing $\alpha$, the peak of $q_N^{\max}(m)$ moves to larger $m$, reflecting the weakening of the small-$x$ trapping associated with the singular infinite invariant density near the marginal fixed point.}
\label{fig:alpha1}
    \end{minipage}
    \hfill
    \begin{minipage}[t]{0.49\linewidth}
        \centering
    \includegraphics[width=\linewidth]{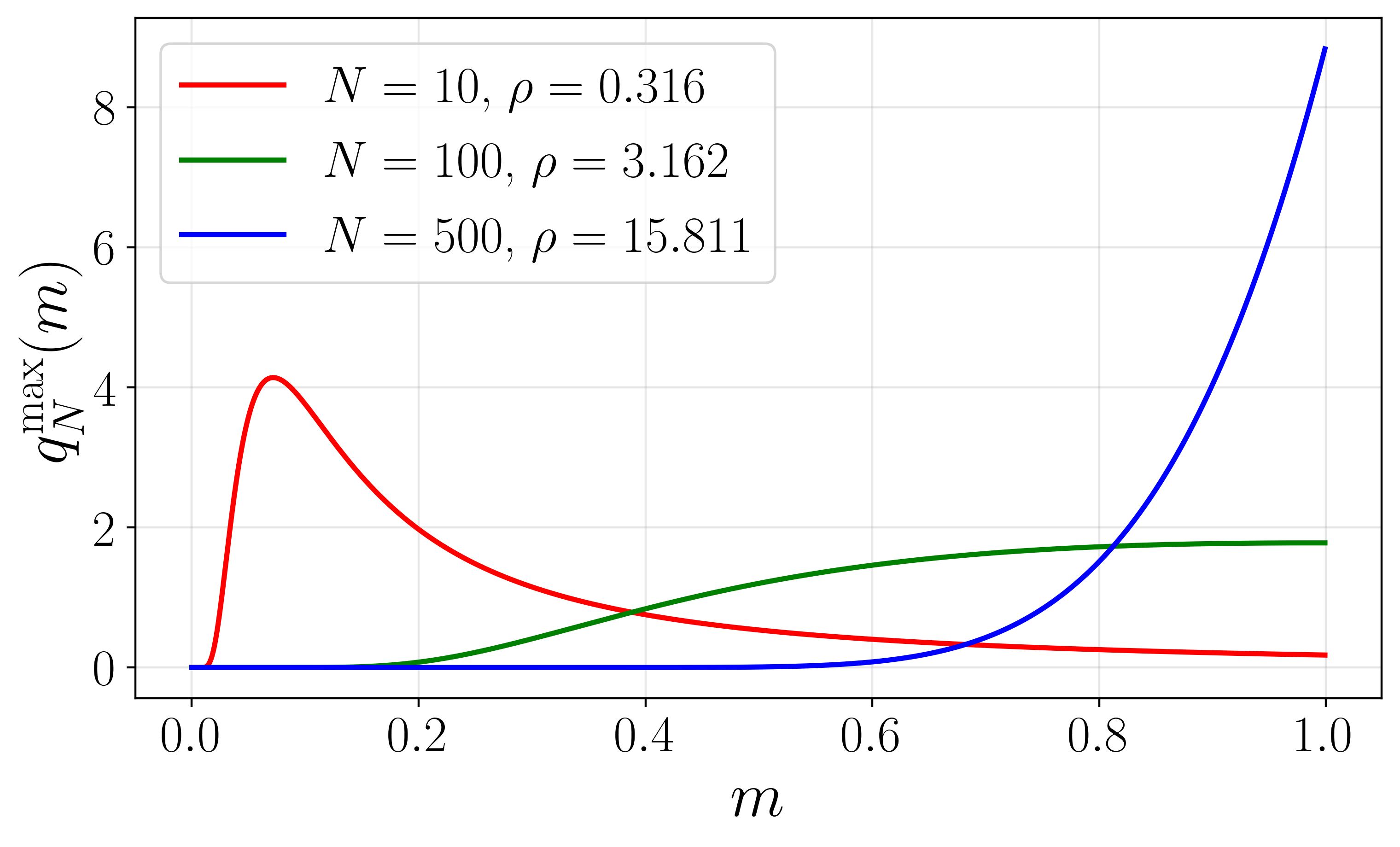}
    \caption{Plot of the solution in Eq.~(\ref{eq:qformap}) for the Thaler map, shown for fixed $\alpha=\tfrac12$ and $t=10^3$, with varying $N$. Here increasing $N$ increases the number of independent opportunities for a rare larger position, so the maximum is progressively pushed toward larger $m$ (closer to $m=1$).
}\label{fig:alpha}
    \end{minipage}
\end{figure}

Figs.~\ref{fig:alpha1} and \ref{fig:alpha} illustrate the parameter dependence of the PDF in Eq.~(\ref{eq:qformap}). As seen in Fig.~\ref{fig:alpha1}, increasing $\alpha$ shifts the distribution toward larger $m$, reflecting the weakening of the small-$x$ trapping. Likewise, in Fig.~\ref{fig:alpha}, at fixed $\alpha=\tfrac12$, increasing $N$ pushes the distribution toward larger $m$, since a larger sample makes rare large-position more likely.

At $m\rightarrow1$,
\begin{equation}\label{eq:qnm1}
    q_N^{max}(m=1)\rightarrow B(1+2^{-1\alpha})\rho \, ,
\end{equation}
so, the boundary value grows linearly with the control parameter $\rho$. The prefactor $B(1+2^{-1/\alpha})$ sets the slope and encodes the $\alpha$-dependence, with larger $\alpha$ corresponding to a larger weight near $m=1$, consistent with weaker trapping near small $x$.

\subsection{Sub-recoil laser cooling}
\subsubsection{The infinite invariant density}
Sub-recoil laser cooling refers to schemes in which atoms are cooled below the single-photon recoil limit by engineering a velocity-selective ``dark'' region in momentum space, so that the scattering (fluorescence) rate is strongly suppressed when the atoms speed is close to zero. This creates and effective trap in velocity space when the velocities are small. Early realizations include velocity-selective coherent population trapping and Raman-based protocols that achieve cooling below the recoil energy scale~\cite{aspect1988laser,kasevich1992laser,davidson1994raman}. A key development was the recognition that these cooling dynamics are naturally described by Lévy statistics and broadly distributed trapping times, leading to non-Gaussian momentum packets and anomalous relaxation~\cite{bardou1994subrecoil,Reichel1995Cs3nK,BardouBook2002,BertinBardou2008AJPhys}. In particular, when the mean trapping time diverges, one expects nonstandard ergodic properties and the emergence of a non-normalizable long-time infinite invariant density description for ensemble and time averages~\cite{barkai2022gas,barkai2021transitions}.

A convenient minimal description is a renewal (jump) process for the speed $v(t)>0$ of the atoms. At $t=0$ one draws a speed $v_1$ from a prescribed reinjection PDF $f(v)$. The speed then remains constant for a random sojourn time $\tilde\tau_1$, whose conditional PDF is exponential with mean lifetime $\tau(v)$,
\begin{equation}\label{eq:qtau}
q(\tilde\tau \,|\, v)=\frac{1}{\tau(v)}\exp\!\left[-\frac{\tilde\tau}{\tau(v)}\right].
\end{equation}
After $\tilde\tau_1$, a new speed $v_2$ is drawn from $f(v)$ and the procedure repeats \cite{barkai2022gas}.

Let $p(v,t)$ denote the PDF of $v$ at time $t$. Under the assumption that the new velocity is drawn independently of the previous one, the rate factorizes as
\begin{equation}
W(v\to v') = R(v)\,f(v'),
\end{equation}
where $R(v)=\frac{1}{\tau(v)}$. This yields a master equation of gain-loss form
\begin{equation}\label{eq:subcoilmaster}
\frac{\partial p(v,t)}{\partial t}
= -\frac{p(v,t)}{\tau(v)} + f(v)\int_0^\infty \frac{p(v',t)}{\tau(v')} \, dv'.
\end{equation}
The steady state solution of the master Eq.\,(\ref{eq:subcoilmaster}) is \begin{equation}\label{eq:ss}
    p_{ss}(v)\propto \tau(v)f(v) \, .
    \end{equation} 
We now need to distinguish between two cases either this steady state can be normalized or not, here we will be interested in the later. In sub-recoil cooling the lifetime diverges algebraically,
\begin{equation}\label{eq:tauv}
\tau(v)\sim c\, v^{-\gamma}, \qquad v\to 0\, .
\end{equation}
From the conditional sojourn-time law at fixed velocity in Eq.\,(\ref{eq:qtau}), the waiting-time PDF between velocity renewals, $\psi(\tilde\tau)$, is obtained by averaging this over the reinjection velocities $\psi(\tilde\tau)=\int_0^{\infty}f(v)q(\tilde\tau,v)dv$. Using Eq.\,(\ref{eq:tauv}) as $v\to0$, the long-time tail is dominated by small velocities and one finds $\psi(\tilde\tau)\sim \tilde\tau^{-1-\alpha}$, we therefore refer to $\alpha$ as the persistence exponent.

For $\gamma>1$ in Eq.\,(\ref{eq:tauv}), the would-be steady state solution is non-normalizable when $f(v) \neq 0$, as $v\rightarrow 0$. Nevertheless, it provides the correct long-time description in the sense of infinite ergodic theory: the rescaled finite-time density converges (for fixed $v>0$) to a non-normalizable {infinite invariant density} $\mathcal{I}(v)$ \cite{bardou2002levy,afek2023colloquium,bertin2008laser,barkai2021transitions,barkai2022gas},
\begin{equation}\label{eq:laserid}
t^{\,1-\alpha}p(v,t)\longrightarrow \mathcal{I}(v),
\qquad \alpha=\frac{1}{\gamma},
\qquad 
\mathcal{I}(v)=\mathcal{B}\,\tau(v)\,\frac{f(v)}{f(0)} \, .
\end{equation}
Hence $\mathcal{N}=1$ in Eq.\,(\ref{eq:infden_def}). Here $\mathcal{B}$ is given by ~\cite{barkai2022gas}
\begin{equation}\label{eq:stupidN}
    \mathcal{B} = \frac{\sin{\pi \alpha}}{\Gamma(1+\alpha)\pi  }c^{-\alpha}\,.
\end{equation}

\begin{figure}[h]
    \centering
    \includegraphics[width=0.5\linewidth]{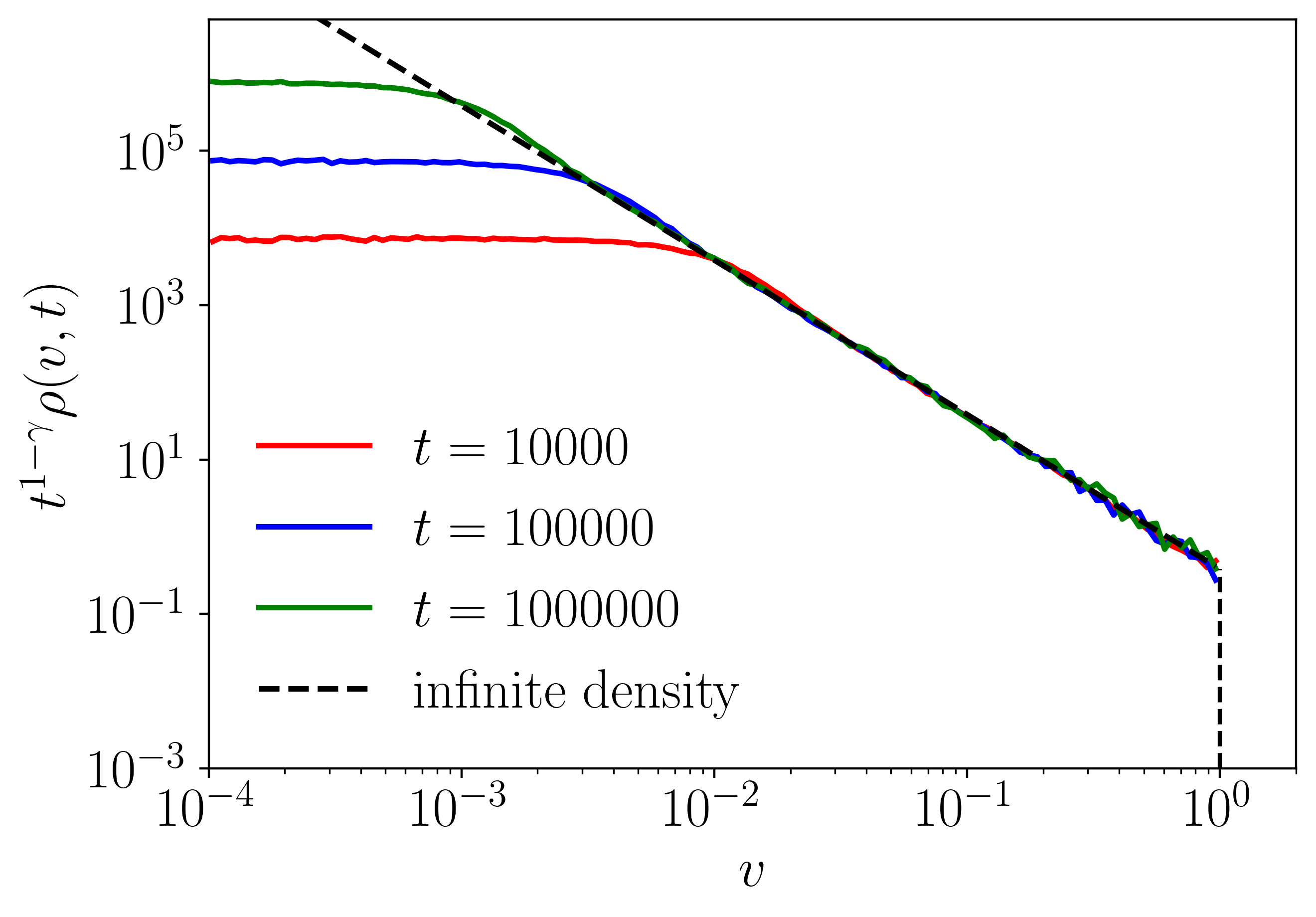}
        \caption{Convergence to the infinite invariant density in sub-recoil cooling. Shown is the rescaled single-particle velocity PDF $t^{1-\alpha}p(v,t)$ with $\alpha=1/\gamma$, obtained from direct simulations at the indicated observation times $t$. We use a direct Monte Carlo renewal simulation: after each reinjection, $v$ is drawn uniformly in $(0,1)$, and the waiting time is drawn from  Eq.\,(\ref{eq:qtau}).
        The rescaled curves collapse onto the infinite invariant density $\mathcal{I}(v)$ (black dashed), given in Eq.\,(\ref{eq:laserid}), with a cutoff at $v=1$. Here we used $\gamma=2$ and $c=1$. The layout follows Fig.\,2 in \cite{barkai2022gas}.
        }
\label{fig:laser_infinite_density}
    \end{figure}

In our numerics we take a uniform parent distribution $f(v)=1$ on $0<v<1$ and an escape rate $R(v)=v^\gamma/c$, which gives
\begin{equation}\label{eq:simulationrecoilinf}
    \mathcal{I}(v) =\mathcal{B}\,c\,v^{-\gamma}\, , \qquad 0<v<1 .
\end{equation}
We generate trajectories by direct simulation of the renewal dynamics: starting from a velocity drawn from $f(v)$, we draw a sojourn time from an exponential distribution with rate $R(v)$, then redraw the velocity from $f(v)$ and repeat. We record $v(t)$ at a fixed observation time over many independent trajectories to estimate $p(v,t)$. The corresponding collapse of $t^{\,1-\alpha}p(v,t)$ onto $\mathcal{I}(v)$
is shown in Fig.~\ref{fig:laser_infinite_density}.
Since $\gamma>1$, $\mathcal{I}(v)$ is non-normalizable, and therefore we must consider case 1 with $\beta=\gamma$ in Eq.\,(\ref{eq:clasess}). Note that for finite times, $\mathcal{I}(v)$ describes the solution of the master equation for the speed $v$ region $v>v_c(t)$  where $v_c(t)$ decays with time as $t^{-\alpha}$. This is similar to the behavior discussed in the previous subsections. Specifically, the speeds in the region $v<v_c(t)$
are irrelevant for the calculation of the maximum.

\subsubsection{Extreme statistics of the maximum speed}
We consider a snapshot of $N$ atoms at a common observation time $t$. Denoting by $v_1,\ldots,v_N$ their speeds at that instant, we define the sample maximum
\begin{equation}
V_{max}=\max\{v_1,\ldots,v_N\}.
\end{equation}
In the long-time regime the single-particle velocity PDF admits an infinite-invariant-density introduced in Eq.\,(\ref{eq:laserid}). 
Accordingly, the probability $s(v,t)=\Pr(v(t)>v)$ is controlled by $\mathcal{J}(v)$, and the
maximum statistics follow directly from the general extreme-value result
in Eq.~\eqref{eq:fM_beta0}, provided we take the joint limit $N,t\to\infty$ with fixed ratio
\begin{equation}
\rho \equiv N\,t^{\alpha-1} \, , 
\label{eq:rho_laser}
\end{equation}
where $\alpha=1/\gamma$ as given in Eq.\,(\ref{eq:laserid}). Thus, for finite $\rho$, the limiting form of the cumulative distribution function $Q_N^{max}(\cdot)$ and probability distribution function $q_N^{max}(\cdot)$ is obtained from
Eqs.\,(\ref{eq:QN_limit_beta0}) and (\ref{eq:fM_beta0}) by identifying $\mathcal{I}(v)$ and $\mathcal{J}(v)$ with the
single-particle infinite density and its integral in the present model.

In particular, for the choice $f(v)=1$ on $0<v<1$, the single-particle infinite density is given in Eq.\,(\ref{eq:simulationrecoilinf}). For maximum statistics, the relevant integral in Eq.\,(\ref{eq:iid_scaling_tail}) is 
\begin{equation}\label{eq:subrecoilJ}
\mathcal{J}(v)=
\int_v^1 \mathcal{I}(u)\,du
=
\begin{cases}
\dfrac{\mathcal{B}\,c}{\gamma-1}\left(v^{1-\gamma}-1\right), & \gamma\neq 1,\\[8pt]
\mathcal{B}\,c\ln\!\left(\dfrac{1}{v}\right), & \gamma=1.
\end{cases}
 .
\end{equation}
Consequently, at the large $t$ and large $N$ limit at fixed $\rho$, for the uniform $f(v)$ 
\begin{equation}\label{eq:subrecoilq}
q_N^{max}(m)= 
\begin{cases}
\rho\,\mathcal{B}\,c \,m^{-\gamma}\exp\!\left[-\rho\,\dfrac{\mathcal{B}\,c}{\gamma-1}\left(m^{1-\gamma}-1\right)\right] , & \gamma\neq 1,\\[8pt]
\rho\,\mathcal{B}\,c\,m^{-1}\exp\!\left[-\rho\,\mathcal{B}\,c\ln\!\left(\dfrac{1}{m}\right)\right], & \gamma=1.
\end{cases}
\, ,
\end{equation}
for $0<m<1$. We have an essential singularity when $m\rightarrow 0$. Physically, it is highly unlikely that the maximum will be very small, hence $q^{max}_N(m)\rightarrow 0$ as $m\rightarrow 0$. In contrast, as $m\to 1$, we have $q_N^{max}(m) \to \rho \,\mathcal{B} \,c$. In this regime, $q_N^{\max}(m)$ increases with $N$ and decreases with $t$. Specifically, we find that the maximum of the PDF $q_N ^{max}(m)$ namely its peak,  is attained at
\begin{equation}\label{eq:peak}
m^*=\min\!\left[\left(\frac{\rho \mathcal{B} c}{\gamma}\right)^{\frac{1}{\gamma-1}},\,1\right].
\end{equation}
 As expected, the probability of reaching high speed is larger when there are more particles, while at longer times the atoms speed becomes concentrated around small speeds due to laser cooling, so the position
 of the peak in Eq.\,(\ref{eq:peak}) shifts to smaller $m*$ as $rho$ decreases.

\begin{figure}[h]
    \centering
    \begin{minipage}[t]{0.49\linewidth}
        \centering
    \includegraphics[width=\linewidth]{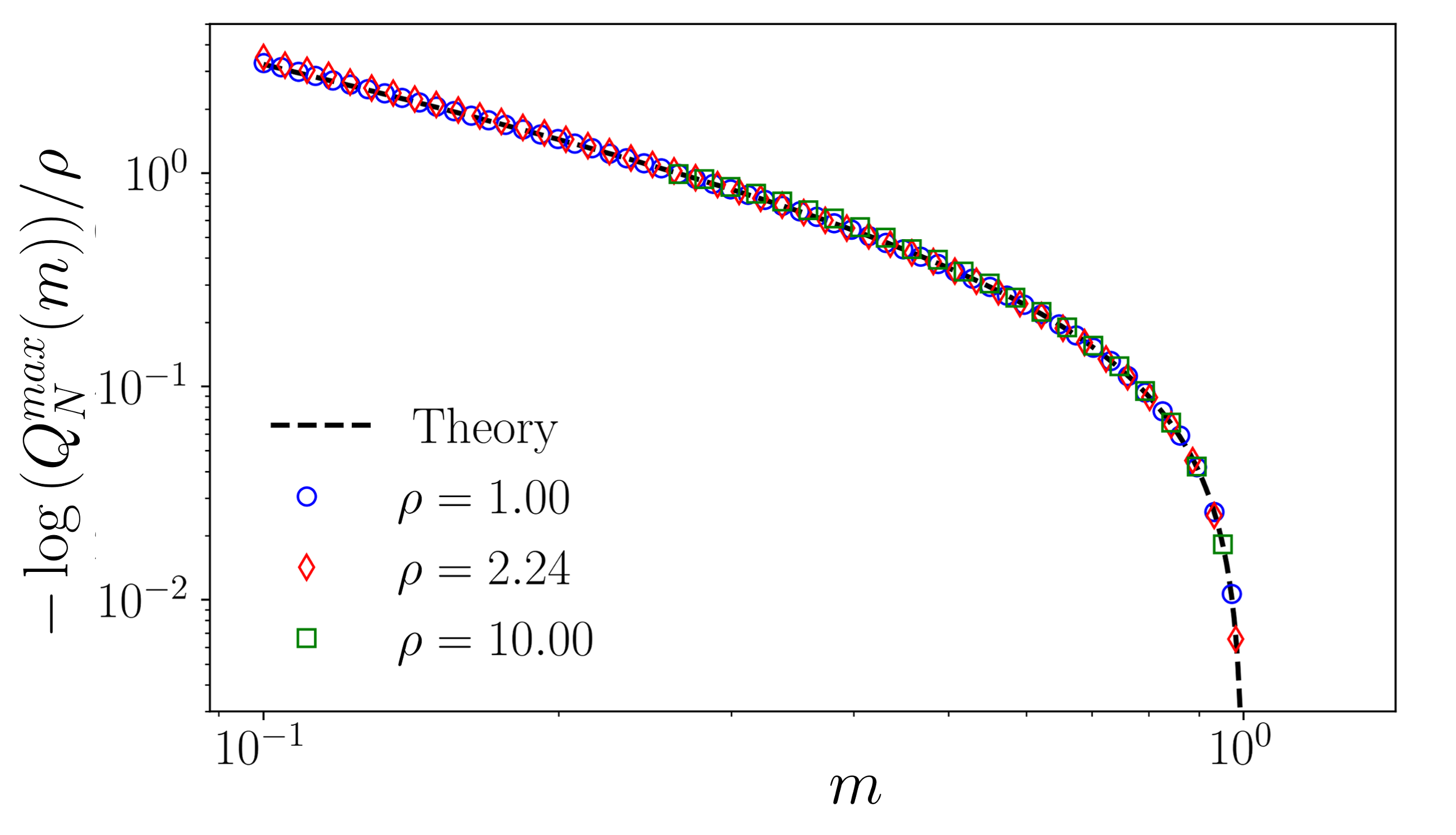}
    \caption{We show extreme-value statistics collapse for the maximum speed of an atom for the laser cooling model. We plot $-\ln Q_N^{max}(m,t)/\rho$, using $N$ independent trajectories at observation time $t$. Curves for different pairs times $t$ for $N=100$, with a fixed ratio $\rho=N\,t^{\alpha-1}$ collapse, confirming the finite-$\rho$ prediction $-\ln Q_N^{max}/\rho\to \mathcal{J}(m)$. Here we used $\gamma=2$.}
\label{fig:laserlog}
    \end{minipage}
    \hfill
    \begin{minipage}[t]{0.49\linewidth}
        \centering
    \includegraphics[width=\linewidth]{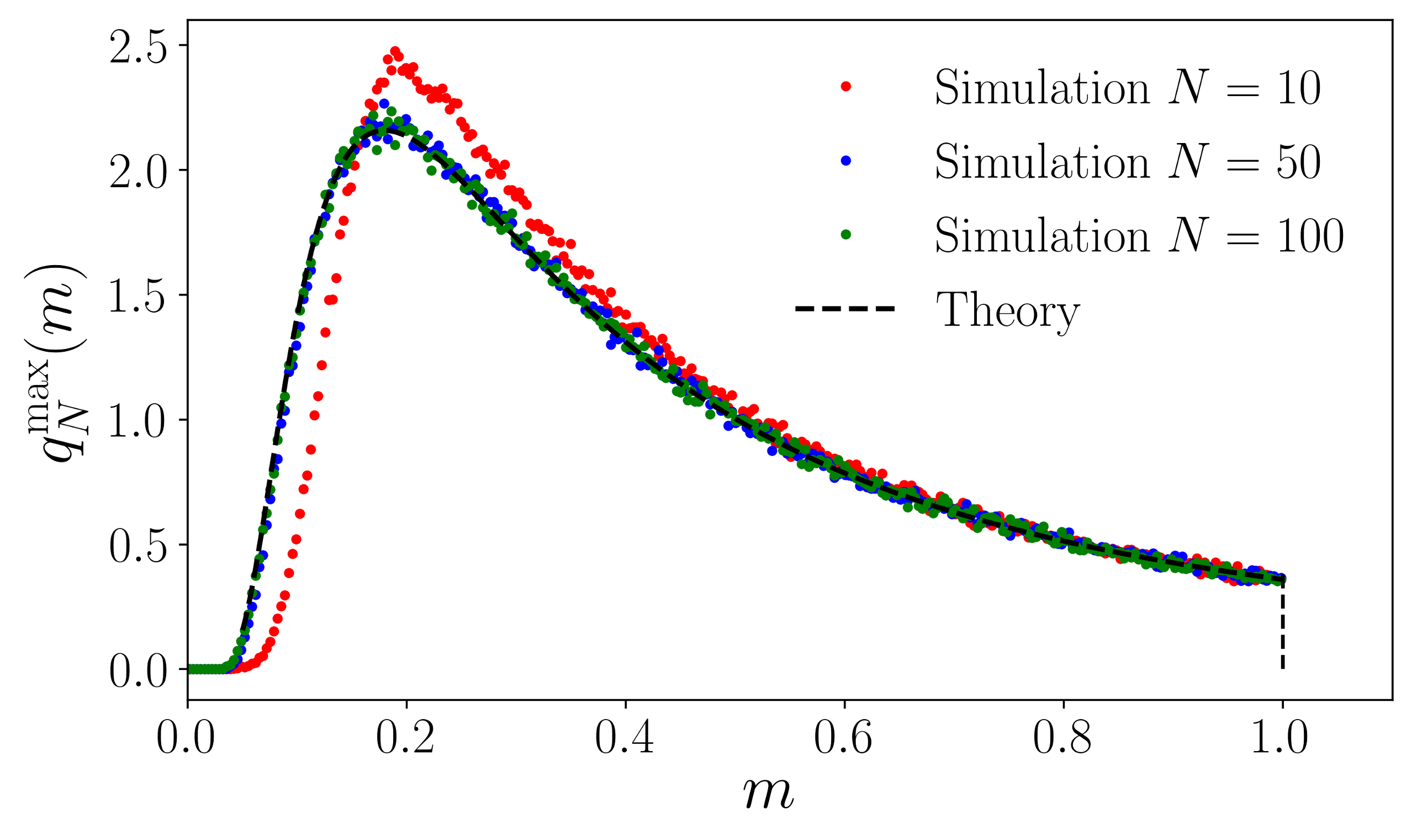}
        \caption{PDF of the maximum speed at fixed $\rho$. We show the probability density $q_N^{max}(m,t)$ of the sample maximum velocity $V_{max}$ obtained from renewal simulations for parameter pairs $(N,t)$ chosen to realize the same ratio $\rho=N\,t^{\alpha-1}$ (here $\rho=2$ and $\gamma=2$). The data converge to the finite-$\rho$ limiting form implied by the general theory in Eq.\,(\ref{eq:subrecoilq}).}
\label{fig:laserqn}
    \end{minipage}
\end{figure}

Numerically, we test the prediction in two complementary ways. In Fig.\,(\ref{fig:laserlog}) we plot $-\ln Q_N^{max}(m,t)/\rho$ versus the maximum $m$ for several values of $\rho$ and observe collapse onto the same integrated infinite density $\mathcal{J}(\cdot)$ in Eq.\,(\ref{eq:subrecoilJ}). Second, in Fig.\,(\ref{fig:laserqn}), for a representative $\rho=2$ we show the corresponding limiting PDF of the maximum, demonstrating convergence toward the prediction in Eq.\,(\ref{eq:subrecoilq}). 
Similarly to the previous examples, changing either the number of particles $N$ or time $t$ alone changes the shape of the extreme-value distribution. In contrast, when $\rho = N/t^{1-\alpha}$ is kept fixed, the distribution shape is preserved when both $N$ and $t$ are large.

\newpage

\section{summary and Discussion}
We formulated extreme value theory for systems whose long time limit is
described by an infinite invariant density. The extreme value statistics is found in a limit where both the number of random variables in the set $N$, and the time $t$ are large, and their ratio defined with $\rho$ remains fixed. The retrun exponent $\alpha$ controls the statistics.
Thus, our theory connects three branches of science, extreme value statistics, infinite ergodic theory, and first passage or persistence theory. There are several open problems.

\begin{itemize}

\item[1.]
The study of order statistics, namely the $k$-th smallest value out of a list of
$N$ random variables, is of interest. As $k$ is varied we may find a transition between
extreme value theory controlled by the infinite density and rare fluctuations,
to a situation where the typical fluctuations of the parent distributions are in control.

\item[2.]
We studied the case where the process $x(t)$ is bounded
from below $x(t)>0$ while for the non-linear PM-like maps it was bounded
from above as well. The structure of the theory for unbounded processes, for example
a particle diffusing in the presence of a finite square well, centered on the origin, is of interest. If particles can
escape this well in both directions  and then at large distances the process is diffusive,
neither the maximum nor the minimum will be related to the non-normalized
Boltzmann measure. However, possibly the statistics of the center particle position would be highly sensitive to the potential shape and the related non-normalized Boltzmann state.

\item[3.] We considered a classification of extreme value theory where the infinite density
is non-normalizable due to its feature at either $x\to 0$ (case 1) or $x \to \infty$ (case 2).
One could have other interesting scenarios, for example maps with more than one marginally unstable
fixed point.

\item[4.] As mentioned, we considered the limit where both $N$ and $t$ are
large while the ratio $\rho$ is fixed. In the "dense" limit,  when
 $N$ is taken to be large while $t$ is fixed, we expect
to find standard extreme value theory. On the other hand, in the sparse limit (briefly studied in Appendix \ref{appen:sparse}), when $N$ is not too large, the sampling of the rare events is reduced, and the role of the infinite density is expected to diminish.
How to transition between the low, intermediate and large $\rho$ limit, is left for future work.

\item[5.] Other open questions remain for the specific model. For example we found
that at low temperature the distribution of the minimum for a particle in
a Lennard Jones potential has a peak (see Fig. \ref{fig:boltzmannqn}). The location of this peak, its departure
from the minimum of potential (as we increase $N$ or $t$ or temperature $T$) is of interest.
The presence of this peak is found since the particle cannot sample the diverging
potential close to the origin. How is this peak and behavior of the extreme value theory
related to the shape of the potential, was explored here, but only briefly.
Similarly, rich shapes of the PDF of the extremum are found, that are controlled by $t, N$ and $z$ (for the maps) or $\gamma$ (for the laser cooling system).

\item[6.] For the non linear maps the return exponent $\alpha$ is controlled by the non linearity $z$ of the map. Similarly, in the laser cooling model, the return exponent is determined by the exponent $\gamma $ describing the rate in the $v\rightarrow 0$ limit. In the model of Langevin diffusion in an asymptotically flat potential, on the other hand, $\alpha=1/2$ is determined by normal diffusion. However, for anomalous diffusion models, such as fractional Brownian motion, random walk in disordered media etc., the exponent $\alpha$ can differ from $1/2$.

\item[7.] In many models an infinite covariant density describes the rare fluctuations instead
of the invariant density under study here, for example, for L\'evy walks \cite{kessler2010infinite,vezzani2019single,rebenshtok2014infinite}. The extreme value theory in this case, will be controlled by the statistics of these events, yet the details must be studied
further. What is however clear, is that the basic approach  under study here is very general.

\end{itemize}

\section{ACKNOWLEDGMENTS}
We would like to thank the Israel Science Foundation for its support through Grant No. 2311/25.

\newpage
\appendix
\section*{Appendices}
\section{Frobenius--Perron Operator}\label{appendix:FP}

Let \(M:[0,1]\to[0,1]\) be a (piecewise) smooth map and \(\psi(x,t)\) the density of an ensemble of points under \(t\) iterations of \(M\).  The {Frobenius–Perron operator} \(\mathcal P\) evolves densities via
\[
  \bigl[\mathcal P\psi\bigr](y)
  \;=\;\sum_{x\,:\,M(x)=y}
    \frac{\psi(x)}{\bigl|M'(x)\bigr|}\,.
\]
In particular, after \(k\) iterations one has
\[
  \psi(x,k)\;=\;\mathcal P^k[\psi_0](x),
\]
where \(\psi_0(x)\) is the initial density (here uniform).

\vspace{1ex}
\noindent\textbf{Application to the Thaler map.}
For the Thaler map
\[
  M(x) \;=\; \frac{x}{\,1 - x + \frac{x}{1+x}\,},
\]
one computes the two inverse branches
\[
  x_{a,b}(y)
  = \frac{y - 1 \pm \sqrt{5y^2 + 2y +1}}{2(y+1)},
  \quad
  x_{b}(y)
  = \frac{y + \sqrt{5y^2 +12y +8}}{2(y+2)},
\]
and its derivative
\(\;M'(x)=(1+2x+2x^2)/(1+x - x^2)^2\).
Numerically, we
\begin{enumerate}
  \item Discretise \(x\in[10^{-6},1-10^{-6}]\) on a uniform grid.
  \item Represent the current density \(\psi(x)\) by its values on this grid.
  \item Build \(\psi_{k+1}(y)\) via cubic interpolation of \(\psi_k\) at the preimages \(x_{a,b}(y)\), weighted by \(\lvert M'(x_{a,b})\rvert^{-1}\).
  \item Renormalise \(\psi_{k+1}\) so that \(\int_0^1\psi_{k+1}(x)\,dx=1\).
\end{enumerate}
Iterating this procedure \(k\) times yields \(\psi(x,k)\).  Finally, to extract the infinite‐time scaling density
\(\displaystyle\bar{\rho}(x)=\lim_{t\to\infty}t^{1-\alpha}\psi(x,t)\) (with \(\alpha=\tfrac12\)), we multiply the \(k\)th iterate by \(k^{1-\alpha}\) and observe convergence as \(k\to\infty\).


\section{Small-$N$ limit}\label{appen:sparse}

All extreme-value results derived above are obtained in a joint long-time/large-sample limit taken at fixed $\rho$ given in Eq.\,(\ref{eq:rho_def_beta0}), as explained above. For any finite $\rho$, while both $N,t$ are large, the extreme is controlled by the infinite invariant density $\mathcal{I}(\cdot)$ and yields a nontrivial limiting law.

By contrast, a small-sample (small $N$) regime, where $t$ grows parametrically faster than $N$ (in the manner specified below), is a singular limit: with high probability the sample does not probe the outer region $x=O(1)$ (i.e., rare large excursions away from the marginal fixed point) that underlies the fixed-$\rho$ extreme statistics, and the extreme is instead determined by a second, regular finite-time scaling function (a normalized “core” scaling function). We collect these special limits below.

\paragraph{Weakly chaotic map.}:
 The fixed-$\rho$ extreme-value scaling is controlled by the infinite invariant density $\mathcal{I}(x)$. In the small-sampling regime $t^{1-\alpha}\gg N$, the maximum typically remains in the laminar core near the marginal fixed point, $x=O(x_c(t))$ with $x_c(t)\sim \alpha^{\alpha}t^{-\alpha}$, and the relevant description is the core scaling form $p(x,t)\sim t^{\alpha}g(x t^{\alpha})$. In particular, for small $x$ exhibits the behavior $p(x,t)\simeq (1/x_c(t))\,q\!\left(x/x_c(t)\right)$ \cite{akimoto2013aging} where
 \begin{equation}
q(u)=\frac{\sin(\pi\alpha)}{\pi\alpha}\,\frac{1}{1+u^{1/\alpha}} .
\label{eq:core_Dynkin}
\end{equation}
namely
\begin{equation}
Q_N^{max}(m,t)=\Bigg[\int_{0}^{m} p(u,t)\,du\Bigg]^N
\simeq
\left[\frac{1}{x_c(t)}\int_{0}^{m} q\!\left(\frac{u}{x_c(t)}\right)\,du\right]^N  \,.
\end{equation}
This scaling describes the dominant behavior on the core scale $m\sim x_c(t)$. For larger $m$ (the outer region, $x=O(1)$), the distribution crosses over to the rare-event sector associated with the infinite-density tail, but in the small-$N$ limit these events are asymptotically too rare to determine the typical extreme statistics.

\begin{figure}[h]
        \centering
\includegraphics[width=0.6\linewidth]{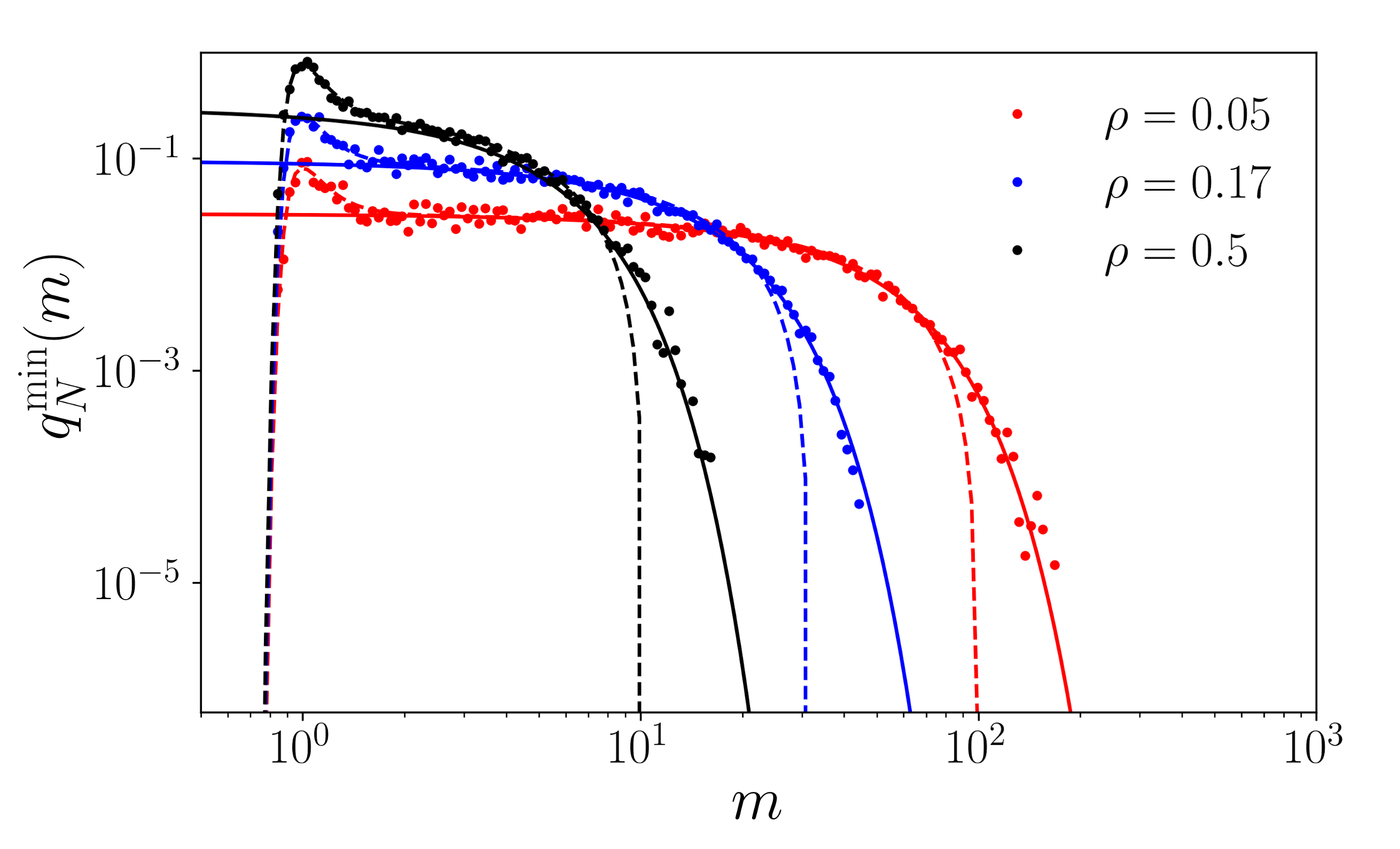}
    \caption{
Crossover in the sparse-sampling limit for overdamped diffusion in an asymptotically flat potential.
Shown is the minimum CDF $Q_N^{min}(m,t)$ (symbols) for
$N=3$ and different values of $\rho$ simulated for $10^5$ trajectories.
For $x$ on the diffusive scale $x=O(\sqrt{Dt})$ (right part of the plot), the data approach the free-diffusion
prediction in Eq.~(\ref{eq:erfqn}) (solid line), reflecting that in the dilute limit it is likely that none of the $N$ trajectories remains in the
potential-dominated region.
Only the small $m$ retains sensitive to the potential and follows the infinite-density-controlled
extreme-value form in Eq.~(\ref{eq:boltzmannqn}) (dashed line).
}
\label{fig:Boltzmann_infinite_2ndscal}
\end{figure}

\paragraph{Overdamped Langevin particle in an asymptotically flat potential.}
While the large-$N$ scaling yields a nontrivial extreme-value law for any finite $\rho$, the small sample, long-time regime $t\gg N^2$ implies $\rho=N/Z_t\to 0$. Then it becomes likely that none of the $N$ particles remains in the central, potential-dominated core, and the minimum is pushed out to the diffusive scale $x=O(\sqrt{Dt})$ where $V(x)\simeq 0$. Consequently, the extreme is determined in addition by free diffusion and one may approximate
\begin{equation}\label{eq:erfqn}
Q_N^{min}(m,t)\simeq 1-\Big[1-\mathrm{erf}\!\Big(\frac{m}{\sqrt{4Dt}}\Big)\Big]^N\, .
\end{equation}
Eq. (\ref{eq:erfqn}) captures the behavior on the diffusive scale and describes the crossover to a “second” (free-diffusive) scaling form at large $m$. Importantly, this does not mean that the potential is irrelevant at all $m$: for sufficiently small $m$ the minimum still probes the rare trajectories that remain in the inner region, and the distribution retains sensitivity to the details of $V(x)$ through the infinite-density-controlled form. This coexistence of a potential-sensitive small-$m$ sector and a free-diffusive large-$m$ sector is illustrated in Fig. \ref{fig:Boltzmann_infinite_2ndscal}.

\paragraph{Sub-recoil laser cooling}:
Beyond the fixed-$\rho$ scaling limit (where the maximum is governed by the infinite-density tail), the sparse-sampling regime $\rho\to 0$ is again singular: the maximum is no longer controlled by the tail and instead remains within the central cooling region $v=O(t^{-\alpha})$. The relevant description is a second scaling form $\rho(v,t)\sim t^{\alpha} g(vt^{\alpha})$, and the limiting law of the rescaled maximum is determined by $g(\cdot)$ rather than by $\mathcal{I}(\cdot)$ \cite{barkai2021transitions}, with 
\begin{equation}
Q_N^{max}(m,t)\simeq
\left[\int_{0}^{m\, t^{\alpha}} g(y)\,dy\right]^N .
\end{equation}
where 
\begin{equation}
    g(x)=\frac{\mathcal{B}}{\alpha x}\exp\!\left(-\frac{x^{1/\alpha}}{c}\right)
\int_{0}^{x}\exp\!\left(\frac{z^{1/\alpha}}{c}\right)\,dz,
\end{equation}
and $\mathcal{B}$ is given in Eq.\,(\ref{eq:stupidN}). This second scaling function therefore governs the typical maximum on the cooling-core scale, while for larger velocities ($m=O(1)$) one crosses over to the outer rare-event sector controlled by the infinite-density tail; in the fixed-$N$ limit these outer events contribute only to the far-velocity tail and do not set the main extreme-value scaling. Notably, $g$ is fixed solely by the small-$m$ asymptotics $\tau(v)\sim c v^{-\alpha}$ and is therefore insensitive to the details of the reinjection mechanism, making this vanishing-$\rho$ extreme-value limit universal within the sub-recoil class.

\bibliographystyle{unsrt}
\bibliography{evbib}

@article{BertinBardou2008AJPhys,
  title     = {{From} laser cooling to aging: a unified {L{\'e}vy} flight description},
  author    = {E. Bertin and F. Bardou},
  journal   = {Am. J. Phys.},
  volume    = {76},
  pages     = {630},
  year      = {2008}
}

@article{baravi2025thermodynamic,
  title     = {{The} thermodynamic limit of extreme first-passage times},
  author    = {T. Baravi and E. Barkai},
  journal   = {Journal of Statistical Mechanics: Theory and Experiment},
  volume    = {2025},
  number    = {12},
  pages     = {123205},
  year      = {2025},
  publisher = {IOP Publishing}
}

@article{samorodnitsky2004extreme,
  title   = {{Extreme} value theory, ergodic theory and the boundary between short memory and long memory for stationary stable processes},
  author  = {G. Samorodnitsky},
  year    = {2004}
}

@article{meyer2017infinite,
  title     = {{Infinite} invariant densities due to intermittency in a nonlinear oscillator},
  author    = {P. Meyer and H. Kantz},
  journal   = {Physical Review E},
  volume    = {96},
  number    = {2},
  pages     = {022217},
  year      = {2017},
  publisher = {American Physical Society},
  doi       = {10.1103/PhysRevE.96.022217}
}

@article{fortin2015applications,
  title     = {{Applications} of extreme value statistics in physics},
  author    = {J.-Y. Fortin and M. Clusel},
  journal   = {Journal of Physics A: Mathematical and Theoretical},
  volume    = {48},
  number    = {18},
  pages     = {183001},
  year      = {2015},
  publisher = {IOP Publishing}
}

@article{embrechts1999modelling,
  title     = {{Modelling} extremal events},
  author    = {P. Embrechts and C. Kl{\"u}ppelberg and T. Mikosch},
  journal   = {British Actuarial Journal},
  volume    = {5},
  number    = {2},
  pages     = {465--465},
  year      = {1999},
  publisher = {Springer}
}

@misc{anderson1984extremes,
  title     = {{Extremes} and related properties of random sequences and processes},
  author    = {C. W. Anderson},
  publisher = {Wiley},
  year      = {1984}
}

@book{Gumbel1958,
  title     = {{Statistics} of extremes},
  author    = {E. J. Gumbel},
  publisher = {Columbia University Press},
  address   = {New York},
  year      = {1958}
}

@article{Gnedenko1943,
  title     = {{Sur} la distribution limite du terme maximum d'une serie aleatoire},
  author    = {B. Gnedenko},
  journal   = {Annals of Mathematics},
  volume    = {44},
  number    = {3},
  pages     = {423--453},
  year      = {1943},
  publisher = {JSTOR}
}

@article{korabel2013numerical,
  title     = {{Numerical} estimate of infinite invariant densities: application to {Pesin}-type identity},
  author    = {N. Korabel and E. Barkai},
  journal   = {Journal of Statistical Mechanics: Theory and Experiment},
  volume    = {2013},
  number    = {08},
  pages     = {P08010},
  year      = {2013},
  publisher = {IOP Publishing}
}

@article{thaler2006distributional,
  title     = {{Distributional} limit theorems in infinite ergodic theory},
  author    = {M. Thaler and R. Zweim{\"u}ller},
  journal   = {Probability Theory and Related Fields},
  volume    = {135},
  number    = {1},
  pages     = {15--52},
  year      = {2006},
  publisher = {Springer}
}

@article{schrodinger1915theorie,
  title   = {{Zur} theorie der fall-und steigversuche an teilchen mit brownscher bewegung},
  author  = {E. Schr{\"o}dinger},
  journal = {Physikalische Zeitschrift},
  volume  = {16},
  pages   = {289--295},
  year    = {1915}
}

@article{rebenshtok2014infinite,
  title     = {{Infinite} densities for {L{\'e}vy} walks},
  author    = {A. Rebenshtok and S. Denisov and P. H{\"a}nggi and E. Barkai},
  journal   = {Physical Review E},
  volume    = {90},
  number    = {6},
  pages     = {062135},
  year      = {2014},
  publisher = {APS}
}

@article{bertin2008laser,
  title     = {{From} laser cooling to aging: a unified {L{\'e}vy} flight description},
  author    = {E. Bertin and F. Bardou},
  journal   = {American Journal of Physics},
  volume    = {76},
  number    = {7},
  pages     = {630--636},
  year      = {2008},
  publisher = {AIP Publishing}
}

@article{aspect1988laser,
  title     = {{Laser} cooling below the one-photon recoil energy by velocity-selective coherent population trapping},
  author    = {A. Aspect and E. Arimondo and R. Kaiser and N. Vansteenkiste and C. Cohen-Tannoudji},
  journal   = {Physical Review Letters},
  volume    = {61},
  number    = {7},
  pages     = {826},
  year      = {1988},
  publisher = {APS}
}

@article{kasevich1992laser,
  title     = {{Laser} cooling below a photon recoil with three-level atoms},
  author    = {M. Kasevich and S. Chu},
  journal   = {Physical Review Letters},
  volume    = {69},
  number    = {12},
  pages     = {1741},
  year      = {1992},
  publisher = {APS}
}

@article{Reichel1995Cs3nK,
  title   = {{Raman} cooling of {Cesium} below 3 {nK}: new approach inspired by {L{\'e}vy} flight statistics},
  author  = {J. Reichel and F. Bardou and M. Ben Dahan and E. Peik and S. Rand and C. Salomon and C. Cohen-Tannoudji},
  journal = {Physical Review Letters},
  volume  = {75},
  pages   = {4575},
  year    = {1995}
}

@book{BardouBook2002,
  title     = {{L{\'e}vy} statistics and laser cooling: how rare events bring atoms to rest},
  author    = {F. Bardou and J.-P. Bouchaud and A. Aspect and C. Cohen-Tannoudji},
  publisher = {Cambridge University Press},
  year      = {2002}
}

@article{davidson1994raman,
  title     = {{Raman} cooling of atoms in two and three dimensions},
  author    = {N. Davidson and H. J. Lee and M. Kasevich and S. Chu},
  journal   = {Physical Review Letters},
  volume    = {72},
  number    = {20},
  pages     = {3158},
  year      = {1994},
  publisher = {APS}
}

@article{bardou1994subrecoil,
  title     = {{Subrecoil} laser cooling and {L{\'e}vy} flights},
  author    = {F. Bardou and J.-P. Bouchaud and O. Emile and A. Aspect and C. Cohen-Tannoudji},
  journal   = {Physical Review Letters},
  volume    = {72},
  number    = {2},
  pages     = {203},
  year      = {1994},
  publisher = {APS}
}

@book{bardou2002levy,
  title     = {{L{\'e}vy} statistics and laser cooling: how rare events bring atoms to rest},
  author    = {F. Bardou},
  publisher = {Cambridge University Press},
  year      = {2002}
}

@article{akimoto2013aging,
  title     = {{Aging} generates regular motions in weakly chaotic systems},
  author    = {T. Akimoto and E. Barkai},
  journal   = {Physical Review E},
  volume    = {87},
  number    = {3},
  pages     = {032915},
  year      = {2013},
  publisher = {APS}
}

@article{barkai2022gas,
  title     = {{Gas} of sub-recoiled laser cooled atoms described by infinite ergodic theory},
  author    = {E. Barkai and G. Radons and T. Akimoto},
  journal   = {The Journal of Chemical Physics},
  volume    = {156},
  number    = {4},
  year      = {2022},
  publisher = {AIP Publishing}
}

@article{barkai2021transitions,
  title     = {{Transitions} in the ergodicity of subrecoil-laser-cooled gases},
  author    = {E. Barkai and G. Radons and T. Akimoto},
  journal   = {Physical Review Letters},
  volume    = {127},
  number    = {14},
  pages     = {140605},
  year      = {2021},
  publisher = {APS}
}

@book{aaronson1997introduction,
  title     = {{An} introduction to infinite ergodic theory},
  author    = {J. Aaronson},
  number    = {50},
  publisher = {American Mathematical Soc.},
  year      = {1997}
}

@article{thaler2000asymptotics,
  title   = {{The} asymptotics of the {Perron}--{Frobenius} operator of a class of interval maps preserving infinite measures},
  author  = {M. Thaler},
  journal = {Studia Mathematica},
  volume  = {143},
  number  = {2},
  pages   = {103--119},
  year    = {2000}
}

@article{thaler1983transformations,
  title     = {{Transformations} on [0, 1] with infinite invariant measures},
  author    = {M. Thaler},
  journal   = {Israel Journal of Mathematics},
  volume    = {46},
  number    = {1},
  pages     = {67--96},
  year      = {1983},
  publisher = {Springer}
}

@article{pomeau1980intermittent,
  title     = {{Intermittent} transition to turbulence in dissipative dynamical systems},
  author    = {Y. Pomeau and P. Manneville},
  journal   = {Communications in Mathematical Physics},
  volume    = {74},
  number    = {2},
  pages     = {189--197},
  year      = {1980},
  publisher = {Springer}
}

@article{korabel2009pesin,
  title     = {{Pesin}-type identity for intermittent dynamics with a zero {Lyapunov} exponent},
  author    = {N. Korabel and E. Barkai},
  journal   = {Physical Review Letters},
  volume    = {102},
  number    = {5},
  pages     = {050601},
  year      = {2009},
  publisher = {APS}
}

@article{grebenkov2026fastest,
  title     = {{F}astest first-passage time for multiple searchers with finite speed},
  author    = {D. S. Grebenkov and R. Metzler and G. Oshanin},
  journal   = {arXiv preprint arXiv:2602.15627},
  year      = {2026}
}

@incollection{lawley2024competition,
  title     = {{C}ompetition of many searchers},
  author    = {S. D. Lawley},
  booktitle = {Target Search Problems},
  pages     = {281--303},
  year      = {2024},
  publisher = {Springer}
}

@article{majumdar2002extreme,
  title     = {{E}xtreme value statistics and traveling fronts: {A}pplication to computer science},
  author    = {S. N. Majumdar and P. L. Krapivsky},
  journal   = {Physical Review E},
  volume    = {65},
  number    = {3},
  pages     = {036127},
  year      = {2002},
  publisher = {APS}
}

@article{jenkinson1955frequency,
  title     = {{T}he frequency distribution of the annual maximum (or minimum) values of meteorological elements},
  author    = {A. F. Jenkinson},
  journal   = {Quarterly Journal of the Royal Meteorological Society},
  volume    = {81},
  number    = {348},
  pages     = {158--171},
  year      = {1955},
  publisher = {Wiley}
}

@article{lawley2023slowest,
  title     = {{S}lowest first passage times, redundancy, and menopause timing},
  author    = {S. D. Lawley and J. Johnson},
  journal   = {Journal of Mathematical Biology},
  volume    = {86},
  number    = {6},
  pages     = {90},
  year      = {2023},
  publisher = {Springer}
}

@article{kessler2010infinite,
  title     = {{Infinite} covariant density for diffusion in logarithmic potentials and optical lattices},
  author    = {D. A. Kessler and E. Barkai},
  journal   = {Physical Review Letters},
  volume    = {105},
  number    = {12},
  pages     = {120602},
  year      = {2010},
  publisher = {APS}
}

@article{aghion2019non,
  title     = {{From} non-normalizable {Boltzmann}--{Gibbs} statistics to infinite-ergodic theory},
  author    = {E. Aghion and D. A. Kessler and E. Barkai},
  journal   = {Physical Review Letters},
  volume    = {122},
  number    = {1},
  pages     = {010601},
  year      = {2019},
  publisher = {APS}
}

@article{mikosch2020gumbel,
  title     = {{Gumbel} and {Fr{\'e}chet} convergence of the maxima of independent random walks},
  author    = {T. Mikosch and J. Yslas},
  journal   = {Advances in Applied Probability},
  volume    = {52},
  number    = {1},
  pages     = {213--236},
  year      = {2020},
  publisher = {Cambridge University Press}
}

@article{zarfaty2022discrete,
  title     = {{Discrete} sampling of extreme events modifies their statistics},
  author    = {L. Zarfaty and E. Barkai and D. A. Kessler},
  journal   = {Physical Review Letters},
  volume    = {129},
  number    = {9},
  pages     = {094101},
  year      = {2022},
  publisher = {APS}
}

@article{oshanin2013anomalous,
  title     = {{Anomalous} fluctuations of currents in {Sinai}-type random chains with strongly correlated disorder},
  author    = {G. Oshanin and A. Rosso and G. Schehr},
  journal   = {Physical Review Letters},
  volume    = {110},
  number    = {10},
  pages     = {100602},
  year      = {2013},
  publisher = {APS}
}

@article{hall1979rate,
  title     = {{On} the rate of convergence of normal extremes},
  author    = {P. Hall},
  journal   = {Journal of Applied Probability},
  volume    = {16},
  number    = {2},
  pages     = {433--439},
  year      = {1979},
  publisher = {Cambridge University Press}
}

@article{majumdar2020extreme,
  title     = {{Extreme} value statistics of correlated random variables: a pedagogical review},
  author    = {S. N. Majumdar and A. Pal and G. Schehr},
  journal   = {Physics Reports},
  volume    = {840},
  pages     = {1--32},
  year      = {2020},
  publisher = {Elsevier}
}

@article{aghion2020infinite,
  title     = {{Infinite} ergodic theory meets {Boltzmann} statistics},
  author    = {E. Aghion and D. A. Kessler and E. Barkai},
  journal   = {Chaos, Solitons \& Fractals},
  volume    = {138},
  pages     = {109890},
  year      = {2020},
  publisher = {Elsevier}
}

@inproceedings{fisher1928limiting,
  title={Limiting forms of the frequency distribution of the largest or smallest member of a sample},
  author={R. A. Fisher and L. H. C. Tippett},
  booktitle={Mathematical proceedings of the Cambridge philosophical society},
  volume={24},
  number={2},
  pages={180--190},
  year={1928},
  organization={Cambridge University Press}
}

@article{bouchaud1992weak,
  title     = {{Weak} ergodicity breaking and aging in disordered systems},
  author    = {J.-P. Bouchaud},
  journal   = {Journal de Physique I},
  volume    = {2},
  number    = {9},
  pages     = {1705--1713},
  year      = {1992},
  publisher = {EDP Sciences}
}

@article{barkai2023ergodic,
  title     = {{E}rgodic properties of {B}rownian motion under stochastic resetting},
  author    = {E. Barkai and R. Flaquer-Galmes and V. M{\'e}ndez},
  journal   = {Physical Review E},
  volume    = {108},
  number    = {6},
  pages     = {064102},
  year      = {2023},
  publisher = {APS}
}

@article{radice2020statistics,
  title     = {{S}tatistics of occupation times and connection to local properties of nonhomogeneous random walks},
  author    = {M. Radice, M. Onofri, R. Artuso and G. Pozzoli},
  journal   = {Physical Review E},
  volume    = {101},
  number    = {4},
  pages     = {042103},
  year      = {2020},
  publisher = {APS}
}

@article{brevitt2025does,
  title     = {{D}oes an intermittent dynamical system remain (weakly) chaotic after drilling in a hole?},
  author    = {S. Brevitt and R. Klages},
  journal   = {New Journal of Physics},
  volume    = {27},
  number    = {10},
  pages     = {104603},
  year      = {2025},
  publisher = {IOP Publishing}
}

@article{akimoto2022infinite,
  title     = {{I}nfinite ergodic theory for three heterogeneous stochastic models with application to subrecoil laser cooling},
  author    = {T. Akimoto and E. Barkai and G. Radons},
  journal   = {Physical Review E},
  volume    = {105},
  number    = {6},
  pages     = {064126},
  year      = {2022},
  publisher = {APS}
}

@article{afek2023colloquium,
  title     = {{C}olloquium: {A}nomalous statistics of laser-cooled atoms in dissipative optical lattices},
  author    = {G. Afek and N. Davidson and D. A. Kessler and E. Barkai},
  journal   = {Reviews of Modern Physics},
  volume    = {95},
  number    = {3},
  pages     = {031003},
  year      = {2023},
  publisher = {APS}
}

@article{biroli2023extreme,
  title     = {{E}xtreme statistics and spacing distribution in a {B}rownian gas correlated by resetting},
  author    = {M. Biroli and H. Larralde and S. N. Majumdar and G. Schehr},
  journal   = {Physical Review Letters},
  volume    = {130},
  number    = {20},
  pages     = {207101},
  year      = {2023},
  publisher = {APS}
}

@article{eliazar2019gumbel,
  title     = {{G}umbel central limit theorem for max-min and min-max},
  author    = {I. Eliazar and R. Metzler and S. Reuveni},
  journal   = {Physical Review E},
  volume    = {100},
  number    = {2},
  pages     = {020104},
  year      = {2019},
  publisher = {APS}
}

@article{tung2025passage,
  title     = {{P}assage times of fast inhomogeneous immigration processes},
  author    = {H.-R. Tung and S. D. Lawley},
  journal   = {Chaos: An Interdisciplinary Journal of Nonlinear Science},
  volume    = {35},
  number    = {12},
  year      = {2025},
  publisher = {AIP Publishing}
}

@article{akimoto2008generalized,
  title     = {{G}eneralized arcsine law and stable law in an infinite measure dynamical system},
  author    = {T. Akimoto},
  journal   = {Journal of Statistical Physics},
  volume    = {132},
  number    = {1},
  pages     = {171--186},
  year      = {2008},
  publisher = {Springer}
}

@article{akimoto2015distributional,
  title     = {{D}istributional behavior of time averages of non-{L} 1 observables in one-dimensional intermittent maps with infinite invariant measures},
  author    = {T. Akimoto and S. Shinkai and Y. Aizawa},
  journal   = {Journal of Statistical Physics},
  volume    = {158},
  number    = {2},
  pages     = {476--493},
  year      = {2015},
  publisher = {Springer}
}

@article{meyer2018anomalous,
  title     = {{A}nomalous diffusion and the {M}oses effect in an aging deterministic model},
  author    = {P. G. Meyer and V. Adlakha and H. Kantz and K. E. Bassler},
  journal   = {New Journal of Physics},
  volume    = {20},
  number    = {11},
  pages     = {113033},
  year      = {2018},
  publisher = {IOP Publishing}
}

@article{ellettari2025rare,
  title     = {{R}are events, many searchers, and fast target reaching in a finite domain},
  author    = {E. Ellettari and G. Nasuti and A. Bassanoni and A. Vezzani and R. Burioni},
  journal   = {arXiv preprint arXiv:2507.09452},
  year      = {2025}
}

@article{giordano2026generalized,
  title     = {{G}eneralized {G}eometric {B}rownian motion and the {I}nfinite {E}rgodicity concept},
  author    = {S. Giordano and R. Blossey},
  journal   = {Philosophical Transactions A},
  year      = {2026},
  note      = {arXiv preprint arXiv:2602.15494}
}

@article{giordano2023infinite,
  title     = {{I}nfinite ergodicity in generalized geometric {B}rownian motions with nonlinear drift},
  author    = {S. Giordano and F. Cleri and R. Blossey},
  journal   = {Physical Review E},
  volume    = {107},
  number    = {4},
  pages     = {044111},
  year      = {2023},
  publisher = {APS}
}

@article{schuss2019redundancy,
  title={Redundancy principle and the role of extreme statistics in molecular and cellular biology},
  author={Schuss, Z and Basnayake, K and Holcman, D},
  journal={Physics of life reviews},
  volume={28},
  pages={52--79},
  year={2019},
  publisher={Elsevier}
}

@article{vezzani2019single,
  title     = {{S}ingle-big-jump principle in physical modeling},
  author    = {A. Vezzani and E. Barkai and R. Burioni},
  journal   = {Physical Review E},
  volume    = {100},
  number    = {1},
  pages     = {012108},
  year      = {2019},
  publisher = {APS}
}

@article{farago2021thermodynamics,
  title     = {{T}hermodynamics of a {B}rownian particle in a nonconfining potential},
  author    = {O. Farago},
  journal   = {Physical Review E},
  volume    = {104},
  number    = {1},
  pages     = {014105},
  year      = {2021},
  publisher = {APS}
}

@article{sandev2026anomalous,
  title     = {{A}nomalous diffusion and fluctuations in complex systems and networks},
  author    = {T. Sandev and L. Kocarev and R. Metzler},
  journal   = {Chaos: An Interdisciplinary Journal of Nonlinear Science},
  volume    = {36},
  number    = {1},
  year      = {2026},
  publisher = {AIP Publishing}
}

@article{endo2025noise,
  title     = {{N}oise-induced transitions in random {P}omeau--{M}anneville maps},
  author    = {T. Endo and Y. Sato and H. Takahasi and E. Barkai and T. Akimoto},
  journal   = {Chaos: An Interdisciplinary Journal of Nonlinear Science},
  volume    = {35},
  number    = {9},
  year      = {2025},
  publisher = {AIP Publishing}
}

@article{brevitt2025singularity,
  title     = {{S}ingularity of {L}{\'e}vy walks in the lifted {P}omeau--{M}anneville map},
  author    = {S. Brevitt and A. Schulz and D. Pegler and H. Kantz and R. Klages},
  journal   = {Chaos: An Interdisciplinary Journal of Nonlinear Science},
  volume    = {35},
  number    = {1},
  year      = {2025},
  publisher = {AIP Publishing}
}

@article{gaspard1988sporadicity,
  title     = {{S}poradicity: between periodic and chaotic dynamical behaviors},
  author    = {P. Gaspard and X.-J. Wang},
  journal   = {Proceedings of the National Academy of Sciences},
  volume    = {85},
  number    = {13},
  pages     = {4591--4595},
  year      = {1988}
}

@article{zumofen1993scale,
  title     = {{S}cale-invariant motion in intermittent chaotic systems},
  author    = {G. Zumofen and J. Klafter},
  journal   = {Physical Review E},
  volume    = {47},
  number    = {2},
  pages     = {851},
  year      = {1993},
  publisher = {APS}
}

@article{geisel1984anomalous,
  title     = {{A}nomalous diffusion in intermittent chaotic systems},
  author    = {T. Geisel and S. Thomae},
  journal   = {Physical Review Letters},
  volume    = {52},
  number    = {22},
  pages     = {1936},
  year      = {1984},
  publisher = {APS}
}

\end{document}